\documentclass[twocolumn, aps, prb]{revtex4}
\usepackage{graphicx}
\usepackage{amsmath}
\usepackage{graphics}
\usepackage{epsfig}
\usepackage{yfonts}
\usepackage{amsmath}
\usepackage{amsfonts}
\usepackage{amssymb}
\usepackage{xcolor}
\usepackage{subfig}
\usepackage{float}
\usepackage{cancel}
\allowdisplaybreaks

\begin{document}

\title{Non-Adiabatic Effects of Nuclear Motion in Quantum Transport of Electrons: A Self-Consistent Keldysh-Langevin Study}

\author{Vincent F. Kershaw } 
\author{  Daniel S. Kosov\footnote{E-mail: daniel.kosov@jcu.edu.au}}
\address{College of Science and Engineering, James Cook University, Townsville, QLD, 4811, Australia }

\begin{abstract}
The molecular junction geometry is  modelled in terms of nuclear degrees of freedom that are embedded in a stochastic quantum environment of non-equilibrium electrons. Time-evolution of the molecular geometry is governed via a mean force, a frictional force and a stochastic force, forces arising from many electrons tunnelling across the junction for a given nuclear vibration. Conversely, the current-driven nuclear dynamics feed  back to the electronic current, which can be captured according extended expressions for the current that have explicit dependencies on classical nuclear velocities and accelerations. Current-induced nuclear forces and the non-adiabatic electric current are computed using non-equilibrium Green's functions via a time-scale separation solution of Keldysh-Kadanoff-Baym equations in Wigner space. Applying the theory to molecular junctions demonstrated that non-adiabatic corrections play an important role when nuclear motion is considered non-equilibrium and, in particular, showed that non-equilibrium and equilibrium descriptions of nuclear motion produce significantly different current characteristics. It is observed that non-equilibrium descriptions generally produce heightened conductance profiles relative to the equilibrium descriptions and provide evidence that the effective temperature is an effective measure of the steady-state characteristics. Finally, we observe that non-equilibrium descriptions of nuclear motion can give rise to the Landauer blowtorch effect via the emergence of multi-minima potential energy surfaces in conjunction with non-uniform temperature profiles. The Landauer blowtorch effect and its impact on the current characteristics, waiting times and the Fano factor are explored for an effective adiabatic potential that morphs between a single, double and triple potential as a function of voltage.

\end{abstract}

\maketitle

\newpage
\section{INTRODUCTION}
The transport of electrons through molecular-sized electronic systems leads to an abundance of new phenomena that are of high physical significance \cite{Galperin2007, Cuevas2017}. An important example of which are the various phenomena related to the interaction between current and molecular vibrational degrees of freedom: a process that results in altered current characteristics of nanoscale electronic devices and, conversely, changed molecular geometries relative to their equilibrium configuration. Electrons tunnelling across the electronic device can exchange energy with the molecule as they travel across the junction; molecular vibrational modes can be activated \cite{Cuevas2017, Wilson1955, Hartle2011, Lu2011, Simine2012, Hartle2018, Peskin2019, Nitzan2018}, often leading to extreme conformational changes in the molecular geometry, and possibly bond rupture within the molecule or at the molecule-lead interface \cite{Gelbwaser-Klimovsky2018, Huang2013, Huang2014,Preston2020}. With this, the changing nuclear degrees of freedom of the molecule may influence the transport properties of the device which, in turn, can lead to interesting transport phenomena such as negative differential resistance, Frank-Condon blockade and non-renewal statistics, amongst others \cite{Hartle2011, Galperin2009, Ioffe2008, Galperin2005, Zazunov2006, Solomon2006, Dzhioev2011, Koch2005,kosov17-nonren}.

\

Experiment and theory have made leaps and bounds in obtaining a fundamental understanding of the interplay between electronic and nuclear degrees of freedom and its impact on electronic current. In terms of quantum-mechanical approaches, a common procedure is the use of non-equilibrium Green's functions (NEGF) to construct various kinds of perturbative expansions about the strength of the electron-vibrational coupling \cite{Paulsson2005, Tikhodeev2001, Galperin2004, Galperin2004a, Egger2008, Entin-Wohlman2009, Mitra2004, Mii2003, Martin-Rodero2008, Monreal2009, Albrecht2012, Dong2013, Laakso2014, SeoaneSouto2014,Caroli1972, Ryndyk2006, Dahnovsky2007, Galperin2006, Ryndyk2007, Hartle2008, Wilner2014, Erpenbeck2015, Frederiksen2007, White2012}. These methods have been applied to the cases of weak and strong electron-nuclear couplings and, in some cases, while treating additional complications such as electron-electron interactions. Other approaches have been proposed in literature to provide numerically accurate solution for different transport scenario in molecular junctions with electron-vibration coupling: examples include master equation approaches \cite{McCarthy2003, Braig2003, Braig2004, Koch2004, Koch2005, Koch2005a, Zazunov2006, Koch2006,dzhioev14,dzhioev15,rudge19-telegraph,kosov17-nonren,kosov17-wtd,kosov18, Hartle2011, Mitra2004a, May2002, Schinabeck2016, Agarwalla2015}, Fock-Liouville space superoperators \cite{Dzhioev2014, Dzhioev2015}, path-integral methods \cite{Muhlbacher2008a}, multilayer multiconfiguration time-dependent Hartree theory \cite{Wang2011a, Wang2013, Wilner2013, Wang2016}, and renormalisation group methods \cite{Laakso2014, Jovchev2013a}, among others \cite{Han2006, Leijnse2008,Erpenbeck2019, Schinabeck2018}. All these methods, while providing tremendous insights into the workings of current-induced phenomena, suffer from one critical flaw: they restrict nuclear dynamics  to vibrations about the fixed equilibrium geometry. According to these models, molecules in nanoscale molecular devices cannot  undergo large-scale structural and/or conformational changes: the result is the neglect of important  physical processes, which can have profound effects on the electronic transport properties of the considered devices, especially at the high voltage and strong electron-vibration coupling regimes.

\

An alternative approach to the problem of vibrationally-induced current is to re-envision the electronic coupling with nuclear dynamics according to that of 'electronic friction' acting on classical nuclei \cite{pistolesi08, pistolesi10, fuse, dzhioev11, Bode11, Bode12, catalysis12, galperin15, Dou2018a, Dou2018b, Dou2017a, Dou2018a, Kershaw2017a, Kershaw2018a, Kershaw2019, PhysRevB.98.041405,Todorov12,Preston2020, Chen2019a}. According to this view, the mechanical degrees of freedom of the molecule are modelled in terms of classical  particles that are embedded in a stochastic electronic environment: time-evolution of the nuclear degrees of freedom is governed via a mean force, a frictional force and a stochastic force, forces arising from many electrons tunnelling across the junction for a given nuclear vibration. Conversely, the molecular junction  geometry has an impact on the electronic current, which can be captured according to extended expressions for the current  that have explicit dependencies on classical nuclear velocities \cite{Bode2011a} and  higher-order terms \cite{Kershaw2017a, Kershaw2018a, Kershaw2019}. Such a formulation holds advantages when compared to some of the former theories mentioned before: it allows for easy descriptions of molecular geometries that can undergo radical deformations; furthermore, such a model provides a gateway into the description of non-adiabatic motion effects in the transport properties of molecular junctions as well as current-induced chemical reactions. Of course, these benefits come at the cost of approximating quantum objects as classical, along with the restriction that nuclear motion must be considered 'slow' relative to the 'fast' tunnelling electrons. 

\

This approach, which we will refer to henceforth as a Keldysh-Langevin description, has received renewed interest in recent years, particularly in the context of non-equilibrium electronic transport \cite{pistolesi08,pistolesi10,fuse,dzhioev11,Bode11, Bode12,catalysis12,Todorov12,Preston2020}. In our previous works \cite{Kershaw2017a, Kershaw2018a, Kershaw2019}, we developed extensions to the Meir-Wingreen formula \cite{meir92} to consider terms that are second order in the central-time derivatives and demonstrated the existence of terms that are appreciable and non-vanishing when statistically averaged over nuclear motion. In this paper, we extend this model by replacing the Boltzmann equilibrium description by a Langevin non-equilibrium description to explore how non-equilibrium nuclear dynamics affects the current characteristics of the device. The theory is based on the assumption that the characteristic time-scales associated with conformational changes of the molecule are slow relative to the fast tunnelling electrons, from which we construct a perturbative transport theory. While perturbative, our theory need not assume that molecular vibrations are small and harmonic about the equilibrium geometry, nor does it assume that the coupling between the electronic and nuclear degrees of freedom be considered small. The result of the theory is an extended formula for the electric current, along with a Langevin equation that self-consistently { { } (all Langevin forces are derived within the model and not postulated)} simulates the current driven nuclear dynamics and current. This paper in many ways serves as an extension to our previous work \cite{Kershaw2017a, Kershaw2018a, Kershaw2019}
where nuclear degrees of freedom are no longer assumed to be in equilibrium but rather change dynamically due to stochastic forces exerted by nonequilibrium electrons. 

\

Section (\ref{Theory}) of the paper illustrates the theoretical model: the Hamiltonian, self-energy and Green's functions are defined with a perturbative solution being implemented for the Keldysh-Kadanoff-Baym equations; and a set of Langevin equations and extended Meir-Wingreen formulas are derived. Section (\ref{Applications}) then applies the proposed theory of electron transport to several problems in order to discuss new phenomena arising out of Langevin descriptions of nuclear dynamics. Note that this paper will work with the atomic unit convention of $e = \hbar = m_{e} = 1$. 

\section{THEORY}
\label{Theory}

\subsection{Hamiltonian}
\label{Hamiltonian}
We start with the general set-up of molecular electronics for a nanoscale quantum system coupled to two macroscopic leads, a set-up that is governed by the Hamiltonian: 
\begin{equation}
\hat{\mathcal{H}}(t) = \hat{H}_{M}(t) + \hat{H}_{L} + \hat{H}_{R} + \hat{H}_{LM} + \hat{H}_{MR}.
\label{hamiltonian}
\end{equation}
Here the Hamiltonian $\hat{\mathcal{H}}(t)$ is the total system Hamiltonian and is comprised of a molecular Hamiltonian $\hat{H}_{M}(t)$, the left and right leads Hamiltonians $\hat{H}_{L}$ and $\hat{H}_{R}$, along with the Hamiltonians $\hat{H}_{LM}$ and $\hat{H}_{MR}$ that describe the coupling between the molecule and the left and right leads, respectively. Note that we have made the time parameter $t$ explicit to show which Hamiltonians carry explicit time-dependence. 

\

Our study will consider a molecular Hamiltonian that has explicit time-dependence through a multi-dimensional vector $\mathbf{x}(t)$ which describes the nuclear degrees of freedom associated with the molecular geometry. The Hamiltonian $\hat{H}_{M}(t)$ takes the form:
\begin{equation}
\hat{H}_{M}(t) = \sum_{ij} h_{ij} ( \mathbf{x}(t) ) \hat{d}^{\dag}_{i} \hat{d}_{j}.
\label{molecularhamiltonian}
\end{equation}
As can be seen above, the second-quantisation operators $\hat{d}^{\dag}_{i}$ and $\hat{d}_{j}$ represent the creation and annihilation operators for the quantum single-particle states $i$ and $j$, along with their respective Hamiltonian matrix elements $h_{ij}(\mathbf{x}(t))$.
 
\

The Hamiltonians for the leads are taken in the standard way as macroscopic reservoirs of non-interacting electrons:
\begin{equation}
\hat{H}_{L} + \hat{H}_{R}  = \sum_{k} \epsilon_{k L} \hat{d}^{\dagger}_{k L} \hat{d}_{k L} + \sum_{k} \epsilon_{k R} \hat{d}^{\dagger}_{k R} \hat{d}_{k R}.
\end{equation}
The creation and annihilation operators $\hat{d}^{\dagger}_{k \alpha}$ and $\hat{d}_{k \alpha}$ (for $\alpha \in \{ L, R \}$) create and annihilate an electron in the leads in the single-particle state $k$ in the $\alpha^{th}$ lead. 

\

Finally, we have the system-lead coupling Hamiltonians $\hat{H}_{LM}$ and $\hat{H}_{MR}$ that take the form:
\begin{multline}
\hat{H}_{LM} + \hat{H}_{MR} = \sum_{k, i} \Big( h_{k L, i}  \hat{d}^{\dagger}_{k L} \hat{d}_{i} + \text{h.c.} \Big) \\ + \sum_{k, i} \Big( h_{k R, i}  \hat{d}^{\dagger}_{k R} \hat{d}_{i} + \text{h.c.} \Big).
\end{multline}
The matrix elements $h_{k \alpha, i}$ (and their conjugates) describe the tunnelling amplitudes between lead states $k \alpha$ and the molecular orbitals $i$. 

\subsection{Green's Functions and Self-Energies}
\label{GFandSE}
Given our time-independent coupling constants $h_{k \alpha, i}$ and $h_{i, k \alpha}$, the self-energy depends on relative time only and therefore becomes time-independent in energy space. The retarded and advanced components take the form:
\begin{equation}
\Sigma_{\alpha i j}^R(\omega)= \Delta_{\alpha i j}(\omega)  -\frac{i}{2} \Gamma_{\alpha i j}(\omega)
\end{equation}
and
\begin{equation}
\Sigma_{\alpha i j}^{A} (\omega) = \Big( \Sigma_{\alpha j i}^R(\omega) \Big)^{*}.
\end{equation}
The level-width functions are given by
\begin{equation}
\Gamma_{\alpha ij} (\omega) = 2 \pi \sum_{k} h^{*}_{i, k \alpha} \delta(\omega-\epsilon_{k \alpha}) h_{k \alpha, j},
\end{equation}
where the level-shift functions $\Delta_{\alpha i j}(\omega)$ can be computed from $\Gamma_{\alpha i j}(\omega)$ via Kramers-Kronig relation. The lesser and greater self-energies are computed as
\begin{equation}
 \Sigma_{\alpha i j}^<(\omega)=  i f_{\alpha}(\omega) \Gamma_{\alpha i j}(\omega)
\end{equation}
and
\begin{equation}
 \Sigma_{\alpha i j}^>(\omega)=  -i (1-f_{\alpha}(\omega)) \Gamma_{\alpha i j}(\omega).
\end{equation}
Using these self-energy expressions, one can readily show that the Kadanoff-Baym equations of motion in the Wigner space take the form \cite{Kershaw2017a, Kershaw2018a, Kershaw2019}:
\begin{multline}
\Big[ \omega  +\frac{i}{2}\partial_T  - e^{\frac{1}{2i}\partial_{\omega}^{{\cal G }} \partial_{T}^{h}}h (T) \Big] \widetilde{{\cal G }}^{R/A}(T,\omega) \\
= I +  e^{-\frac{1}{2i}\partial_{\omega}^{\Sigma} \partial_{T}^{{\cal G }}} {\Sigma}^{R/A}(\omega)\widetilde{{\cal G }}^{R/A}(T,\omega)
\end{multline}
and
\begin{multline}
\Big[ \omega  +\frac{i}{2}\partial_T  - e^{\frac{1}{2i}\partial_{\omega}^{{\cal G }} \partial_{T}^{h}}h (T) \Big] \widetilde{{\cal G }}^{</>}(T,\omega) \\
= e^{-\frac{1}{2i}\partial_{\omega}^{\Sigma} \partial_{T}^{{\cal G }}} \Big[ {\Sigma}^{R}(\omega)\widetilde{{\cal G }}^{</>}(T,\omega) + {\Sigma}^{</>}(\omega)\widetilde{{\cal G }}^{A}(T,\omega) \Big].
\end{multline}
Following our previous works \cite{Kershaw2017a, Kershaw2018a, Kershaw2019}, a solution to these equations can be facilitated by assuming that the characteristic timescale of nuclear vibrations $\Omega$ is smaller than the electron tunnelling timescale $\Gamma$. Such an assumption allows us to express all exponentials above in terms of Maclaurin series' and truncate them to the second order in the central-time derivatives and solve the equations of motion according to a general perturbative ansatz:
\begin{equation}
\mathcal{G} = G + \lambda G_{(1)} + \lambda^2 G_{(2)}, 
\end{equation}
where above we have the smallness parameter $\lambda$ and the adiabatic, first and second-order Green's functions $G$, $G_{(1)}$ and $G_{(2)}$, respectively. All components of the adiabatic Green's functions depend instantaneously on the molecular geometry and are defined though standard relations:
\begin{equation}
G^R(\mathbf x ,\omega) = (\omega I - h(\mathbf x) - \Sigma^R(\omega))^{-1},
\end{equation}
\begin{equation}
G^A(\mathbf x ,\omega) = (\omega I - h(\mathbf x) - \Sigma^A(\omega))^{-1},
\end{equation}
\begin{equation}
G^<(\mathbf x ,\omega) = G^R(\mathbf x ,\omega)  \Sigma^<(\omega) G^A(\mathbf x ,\omega)
\end{equation}
and
\begin{equation}
G^>(\mathbf x ,\omega) = G^R(\mathbf x ,\omega)  \Sigma^>(\omega) G^A(\mathbf x ,\omega).
\end{equation}
Above we have the identity matrix $I$ in molecular orbital space. In the interest of presentation, we will not present expressions for the first-order and second-order Green's functions here (see \cite{Kershaw2017a, Kershaw2018a, Kershaw2019} for details). We note only that these corrections contain information on motion corrections to the adiabatic base case and are composed of adiabatic Green's functions. Soon-to-be-defined quantities such as, for example, viscosity, diffusion and the electric current terms are, as a result, composed of adiabatic Green's functions.

\subsection{The Langevin Equation}
The nuclear degrees of freedom are considered as classical variables within our approach. If, additionally, we assume that nuclear dynamics are slow relative to the tunnelling electrons, we can obtain a Langevin equation of motion for each component $\mu$ \cite{pistolesi08,pistolesi10,Bode11, Bode12,catalysis12,Todorov12,Preston2020},
\begin{equation}
\frac{d p_{\mu}}{d t} =  - \partial_\mu U + F_\mu - \sum_\nu \xi_{\mu \nu} \dot x_\mu  + \delta f_{\mu},
\end{equation}
where we have used the notation $\partial_{\mu} \equiv \frac{\partial}{\partial x_{\mu}}$. The Langevin equations above are comprised of a classical and external classical potential $U(\mathbf x)$, an adiabatic drag force $F_\mu$, a frictional force and its viscosity tensor $\xi_{\mu \nu}$, and a stochastic force $\delta f(t)$. The adiabatic drag force $F_\mu$ has the form
\begin{equation}
F_\mu(\mathbf x ) = i \int \frac{d \omega}{2 \pi} \text{Tr} \Big[ \Lambda_\mu G^<(\mathbf x, \omega) \Big],
\end{equation}
with the matrix $ \Lambda_\mu$ being the derivative of single-particle molecular Hamiltonian matrix:
\begin{equation}
\Lambda_\mu = \partial_\mu h.
\end{equation}
The viscosity tensor which depends on the molecular junction geometry and is given by \cite{Bode12}
\begin{multline}
\xi_{\mu \nu} (\mathbf x)= \int \frac{d\omega}{2 \pi} \text{Tr} \Big[ G^<(\mathbf x, \omega)\Lambda_\mu \partial_\omega G^R(\mathbf x, \omega) \Lambda_\mu 
\\
- G^<(\mathbf x, \omega)\Lambda_\nu \partial_\omega G^A(\mathbf x, \omega) \Lambda_\nu 
\Big].
\end{multline}
The stochastic force $\delta f_{\mu}(t)$ is modelled as a Markovian Gaussian variable with mean
\begin{equation}
\langle \delta f_{\mu}(t) \rangle = 0
\end{equation}
and delta-function variance
\begin{equation}
\langle \delta f_{\mu}(t) \delta f_{\nu} (t^{\prime}) \rangle = D_{\mu \nu} \delta (t - t^{\prime}).
\label{RARA}
\end{equation}
Above we have the diffusion coefficient $D_{\mu \nu}$; one can show that $D_{\mu \nu}$ can be expressed in terms of lesser and greater Green's functions \cite{Bode12}:
\begin{equation}
D_{\mu \nu}(\mathbf x) = \int \frac{d \omega}{2 \pi} \text{Tr} \Big\{ \Lambda_{\mu} G^{<}(\mathbf x, \omega) \Lambda_{\nu} G^{>}(\mathbf x, \omega) \Big\}.
\end{equation}

\subsection{Electric Current}
\label{ElectricCurrent}
A derivation of a formula for electric current proceeds routinely by transforming the equations of motion to the Wigner space and expanding the exponential operator to consider only those terms that are second-order in central-time derivatives, an expansion that is performed in the same assumption as in the derivation of the Langevin equation for nuclear degrees of freedom that are slow relative to the tunnelling electrons.\cite{Kershaw2017a, Kershaw2018a, Kershaw2019} While questions of consistency can be raised in regards to only considering corrections to the first order in the Langevin equation and corrections to the second order in the electric current, we note that we do this for pragmatic purposes. Computing the expectation value  of current in the long-time limit results in  vanishing first-order contributions; considering second-order terms provides us with a first non-vanishing contributions to electronic current when averaging.

\

The expression for the electric current can be cast into the form of the Landauer formula \cite{Kershaw2017a} 
\begin{equation}
\mathcal{J}_{\text{pert}}  = \int^{+\infty}_{-\infty} d\omega
\langle \mathcal{T}(\omega) \rangle_{x} (f_L(\omega)-f_R(\omega)),
\label{landauer}
\end{equation}
where $ \langle ... \rangle_x $ means the statistical averaging over the Langevin trajectory. The form of equation \ref{landauer} guarantees preservation of the continuity  equation when considering terms up to the second order; it should be noted, however, that gradient corrections in Kadanoff-baym equations are not always known to satisfy current conservation (see reference \cite{Kershaw2018a}). First-order corrections, for example, are known violate  conservation laws \cite{Ivanov2000} and can be fixed via the Botermans and Malfliet approximation \cite{Botermans1990}. We are not aware of a similar method to fix this problem in the second-order gradient expansion. The subscript "pert" indicates that this is a current based on the perturbative calculations in terms of nuclear velocities. The transmission coefficient $\mathcal{T}(\omega)$ is decomposed into three parts:
\begin{multline}
\langle {\cal T}(\omega) \rangle_x =  \langle T^{(0)}(\omega) \rangle_x + \sum_{\mu \nu} \langle \dot x_\mu \dot x_\nu A_{\mu \nu}(\mathbf x, \omega) \rangle_x \\ + \sum_{\mu} \langle \ddot x_\mu  B_{\mu}(\mathbf x, \omega) \rangle_x.
\label{transmission_corrections}
\end{multline}
The first part $T^{(0)}(\omega)$ is the standard transmission coefficient which instantaneously depends on the molecular geometry:
\begin{equation}
T^{(0)}(\omega) = - \frac{1}{\pi} \text{Tr} \left[\frac{\Gamma_L(\omega) \Gamma_R(\omega)}{\Gamma_L(\omega) + \Gamma_R(\omega)} \; \text{Im} \Big\{
G^R(\mathbf x, \omega) \Big\} \right].
\label{T0}
\end{equation}
The ratio of $\Gamma$-matrices in the expression above (and the dynamical corrections to follow) is well-defined under the assumption that the $\Gamma$-matrices are proportional such that $\Gamma_{L} = \lambda \Gamma_{R}$ \cite{Textbook}. The last two terms are the dynamical corrections to the transmission coefficient due to nuclear motion. These terms have more cumbersome expressions and it would be beneficial for presentation purposes to define the following two quantities (note that we have suppressed function notation) \cite{Kershaw2017a}:
\begin{equation}
\mathcal{A}^{R}(\omega) = I - \partial_{\omega} \Sigma^{R} 
\end{equation}
and
\begin{equation}
\mathcal{C}^R_{\mu}(\mathbf x, \omega) = \frac{1}{2i} G^R \Big[ \mathcal{A}^R G^R, \Lambda_{\mu} G^R \Big]_{-}.
\end{equation}
The coefficients $A_{\mu \nu}(\mathbf x, \omega)$ and $B_{\mu}(\mathbf x, \omega)$ can then be expressed as:
\begin{multline}
A_{\mu \nu} (\mathbf x, \omega) = - \frac{1}{\pi} \text{Tr} \Big[\frac{\Gamma_L \Gamma_R}{\Gamma_L + \Gamma_R} \text{Im} 
\Big\{ \frac{1}{2i} G^R \mathcal{A}^R \partial_{\nu} \mathcal{C}^R_{\mu} \\ + \frac{1}{2i} G^R \Lambda_{\mu} \partial_{\omega} \mathcal{C}^R_{\nu} + \frac{1}{8} G^R \Phi_{\mu \nu} \partial^2_{\omega} G^R + \frac{1}{8} G^R \partial^2_{\omega} \Sigma^{R} \partial_{\mu \nu} G^R \\ - \frac{1}{8} \partial^2_{\omega} \partial_{\mu \nu} G^R \Big\} \Big]
\end{multline}
and
\begin{multline}
B_{\mu} (\mathbf x, \omega) = - \frac{1}{\pi} \text{Tr} \Big[ \frac{\Gamma_L \Gamma_R}{\Gamma_L + \Gamma_R} \text{Im} \Big\{ \frac{1}{2i} G^R \mathcal{A}^R \mathcal{C}^R \\ + \frac{1}{8} G^R \Lambda_{\mu} \partial^2_{\omega} G^R + \frac{1}{8} G^R \partial^2_{\omega} \Sigma^{R} \partial_{\mu} G^R - \frac{1}{8} \partial_{\mu} \partial^2_{\omega} G^R \Big\} \Big],
\end{multline}
where we have omitted functional dependence on $\mathbf x$ and $\omega$ in the right sides of the equations for brevity. 

\section{Applications}
\label{Applications}
The  general theory presented in section \ref{Theory} is now applied to several transport scenarios. The model system consists of a single molecular orbital coupled to a single nuclear degree of freedom. We structure this section as follows: first, we detail the model based expressions for viscosity, diffusion, and electric current and we also discuss the semi-analytical exact solution  for the considered model. Next, we benchmark the approximate current computed within our approach against the exact results; and, finally, we consider the existence of the blowtorch effect and dynamical blockade of electric current, respectively. Note that all numerical values in the text and figures are given in atomic units.

\subsection{Model}
Before applying the theory to several model problems, we first simplify some of the general equations above in the limit of a single electronic energy level and a single nuclear DOF. We model the molecular-orbital energy as being dependent on a single classical and Cartesian DOF $x(t)$, with the corresponding Hamiltonian being given by 
\begin{equation}
\hat{H}_{M}(t) = h(x(t)) \hat{d}^{\dagger} \hat{d},
\end{equation}
where we have neglected electronic spin. Furthermore, we model the Hamiltonian's dependency on nuclear position in a linear fashion such that:
\begin{equation}
h(x(t)) = h_{0} + \lambda x(t),
\label{Hami}
\end{equation}
where $\lambda$ is a strength of electronic-nuclear coupling and $h_{0}$ is a fixed component of the energy level. We also choose to work in the wide-band approximation where the level-broadening functions for the left and right leads ($\Gamma_{L}$ and $\Gamma_{R}$, respectively) are energy-independent constants. The practical result is that 
\begin{equation}
\partial_{\omega} \Sigma^{A}_{L} = \partial_{\omega} \Sigma^{A}_{R} = 0,
\end{equation}
and hence they disappear from all expressions. Furthermore, the limit of a single electronic energy level means that all fundamental quantities - the Hamiltonian and its derivatives, the Green's functions and self-energies - become numbers rather than matrices and this simplifies the expressions considerably.

\

In these limits, the diffusion coefficient, the frictional coefficient and the adiabatic force are given by:
\begin{equation}
F = - \frac{\lambda}{2 \pi} \int d \omega \frac{\Gamma_{L} f_{L} + \Gamma_{R} f_{R}}{(\omega - h(x))^2 + \Gamma^2 / 4},
\label{adiabaticforces}
\end{equation}
\begin{multline}
\xi = \frac{\lambda^2}{4 \pi} \frac{\Gamma}{kT} \int d \omega \frac{\Gamma_{L} f_L (1 - f_L) + \Gamma_{R} f_R (1 - f_R)}{\big[ (\omega - h(x))^2 + \Gamma^2 / 4 \big]^2}
\end{multline}
and 
\begin{multline}
D = \frac{\lambda^2}{2 \pi} \int d \omega \frac{ \big( \Gamma^2_{L} + \Gamma_{L} \Gamma_{R} \big) f_{L} + \big( \Gamma^2_{R} + \Gamma_{L} \Gamma_{R} \big) f_{R}}{\big[ (\omega - h(x))^2 + \Gamma^2 / 4 \big]^2} \\ - \frac{\lambda^2}{2 \pi} \int d \omega \frac{ \big( \Gamma_{L} f_L + \Gamma_{R} f_{R} \big)^2}{\big[ (\omega - h(x))^2 + \Gamma^2 / 4 \big]^2}.
\end{multline}
The coefficients $A$ and $B$ which determine the dynamic corrections to electric current become
\begin{equation}
A =  - \frac{3 \lambda^2 \Gamma_{L} \Gamma_{R} }{32 \pi} \frac{ (\Gamma^4 
- 40 \Gamma^2(\omega-h(x))^2 + 80 (\omega-h(x))^4}{ (\omega-h(x))^2 + \Gamma^2/4}
\label{eqnfirst}
\end{equation}
and 
\begin{equation}
B =  \frac{\lambda \Gamma_{L} \Gamma_{R} (\omega-h(x))}{8\pi} \frac{\Gamma^2 -4(\omega -h(x))^2}{ (\omega-h(x))^2 + \Gamma^2/4}.
\label{eqnsecond}
\end{equation}
This model also enables us to find the exact numerical solution of the model. As shown in the appendix, we can derive the following expression for the transmission coefficient
\begin{multline}
\big\langle {T}(\omega) \big\rangle_{x} = \frac{1}{\pi} \frac{ \Gamma_{L} \Gamma_{R}}{\Gamma_{L} + \Gamma_{R}} \int^{t}_{- \infty} dt_{1} \\ \times \text{Re} \Big\{ e^{ - i \big( \omega - i \frac{\Gamma}{2} \big) (t - t_{1})} \Big\langle e^{- i \int^{t}_{t_{1}} d t_{2} h (x(t_{2}))} \Big\rangle_{x} \Big\},
\label{transmission}
\end{multline}
to compute the stochastically-averaged current from the left lead according to
\begin{equation}
\mathcal{J}_{\text{exact}} = \int^{\infty}_{-\infty} d \omega \big\langle {T(\omega)} \big\rangle_{x} \Big( f_{L} - f_{R} \Big).
\label{current}
\end{equation}
The current equation (\ref{current}) and the transmission coefficient in (\ref{transmission}) serve as the expressions for the analytical current. Given knowledge of $h(t)$ for a range of stochastic trajectories, then it follows that one can compute exact current trajectories for benchmarking purposes. Subsequent discussions will frequently make use of the short-hand notation $\mathcal{J}_{\text{exact}}$ when referring to the analytical current.

\subsection{Assessment of the Perturbative Expansion}
\label{Electric Current Benchmarking}
\begin{figure*}
    \centering 
    \subfloat[]{{\includegraphics[trim=95 0 95 0, clip, width=8.5cm]{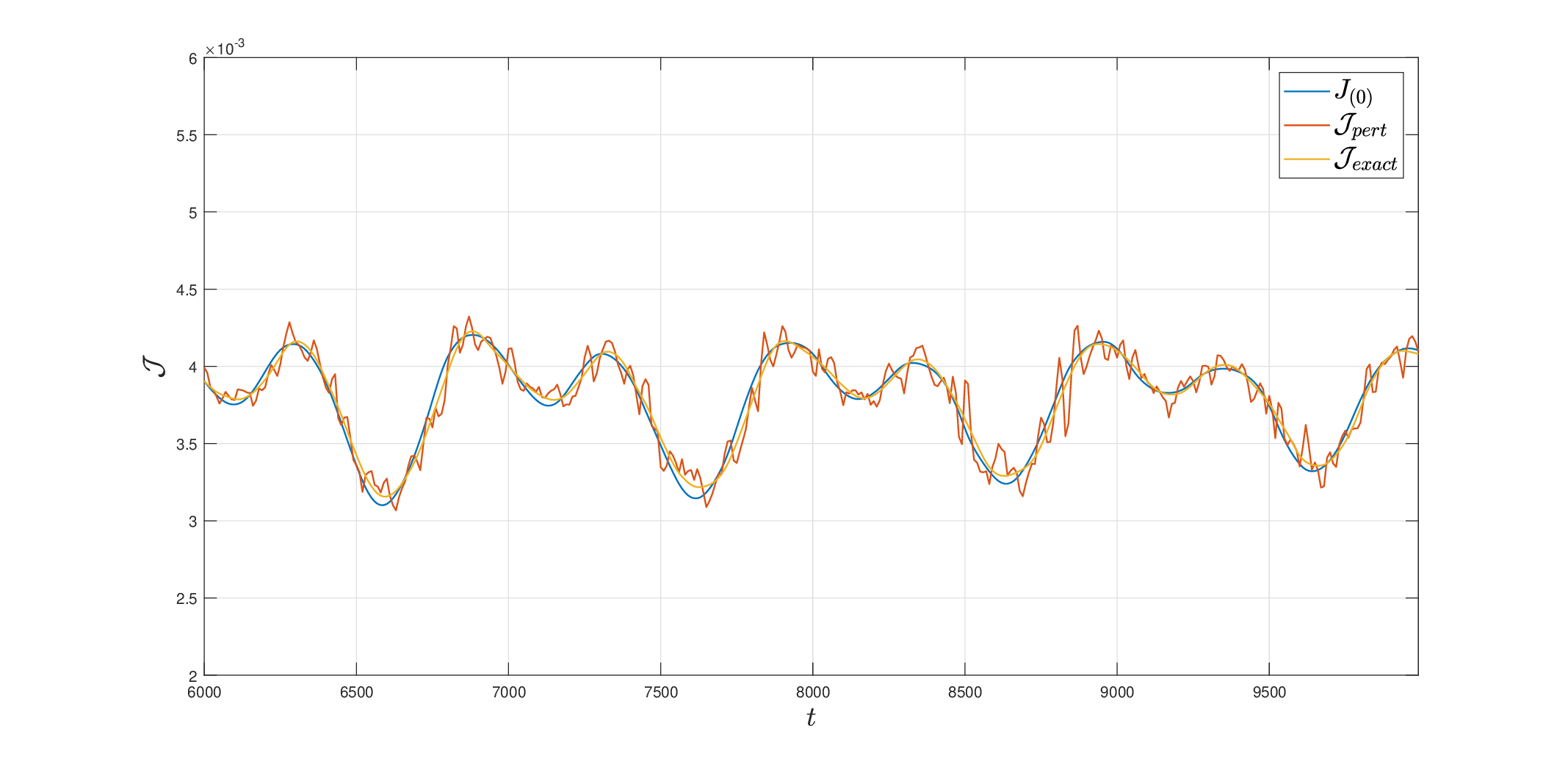} }}%
    \qquad
    \subfloat[]{{\includegraphics[trim=95 0 90 5, clip, width=8.5cm]{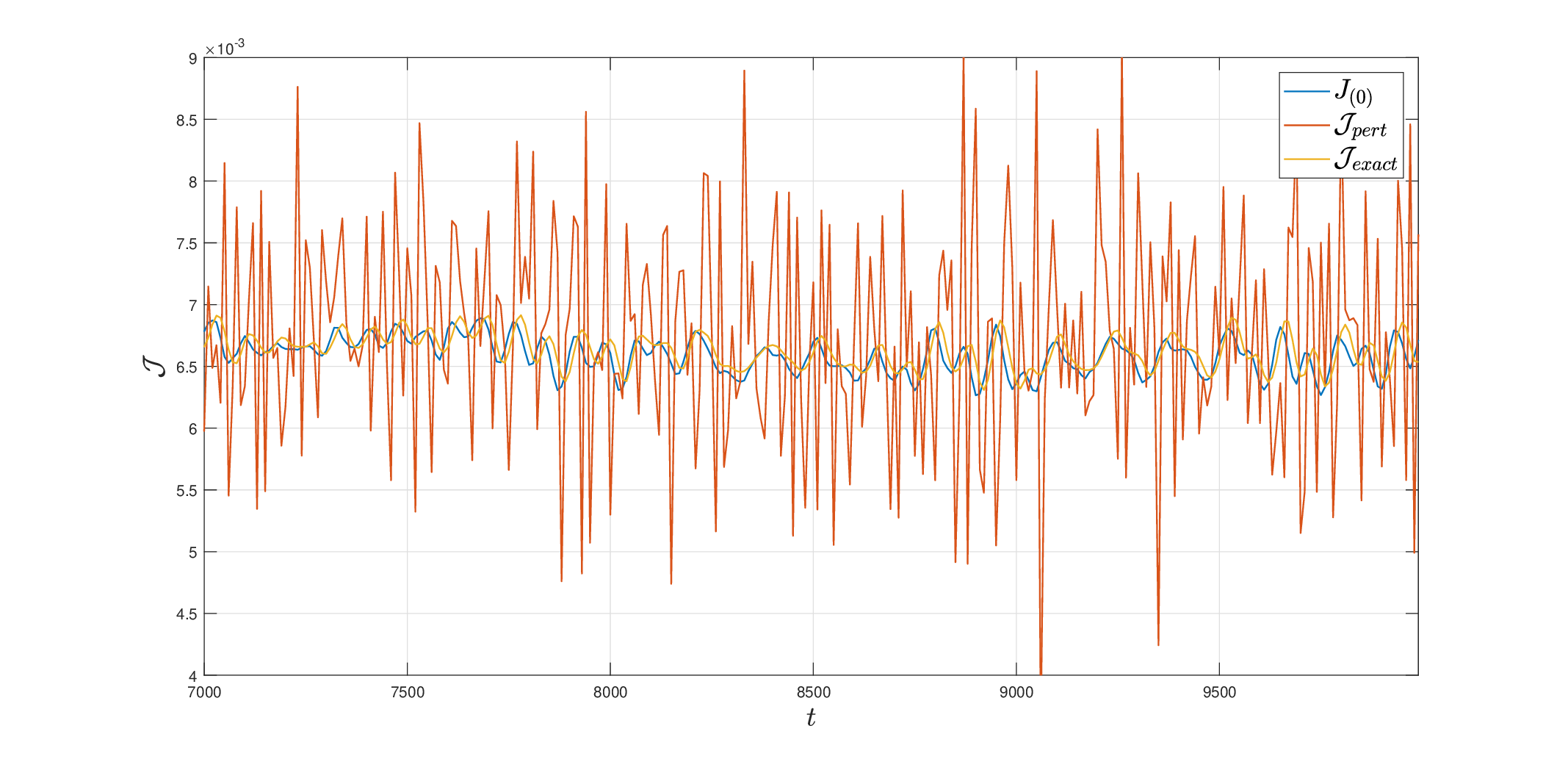} }}%
    \caption{Comparison of $\mathcal{J}_{\text{pert}}$, $\mathcal{J}_{\text{exact}}$ and $J_{(0)}$. Simulation is performed for the electronic parameters of $\Gamma_{L} = \Gamma_{R} = 0.01$, $\lambda = 0.05$, $k = 0.1$ and $V = 0.1$.  In (a) we have the nuclear parameters of $m = 5000$ and (b) considers $m = 1000$. This gives a characteristic frequency of $\Omega = 0.0045$ for a) and $\Omega = 0.01$ for b).}
    \label{TrajectoriesComparison}%
\end{figure*}

\begin{figure*}
    \centering 
    \subfloat[]{{\includegraphics[trim=95 0 95 0, clip, width=8.5cm]{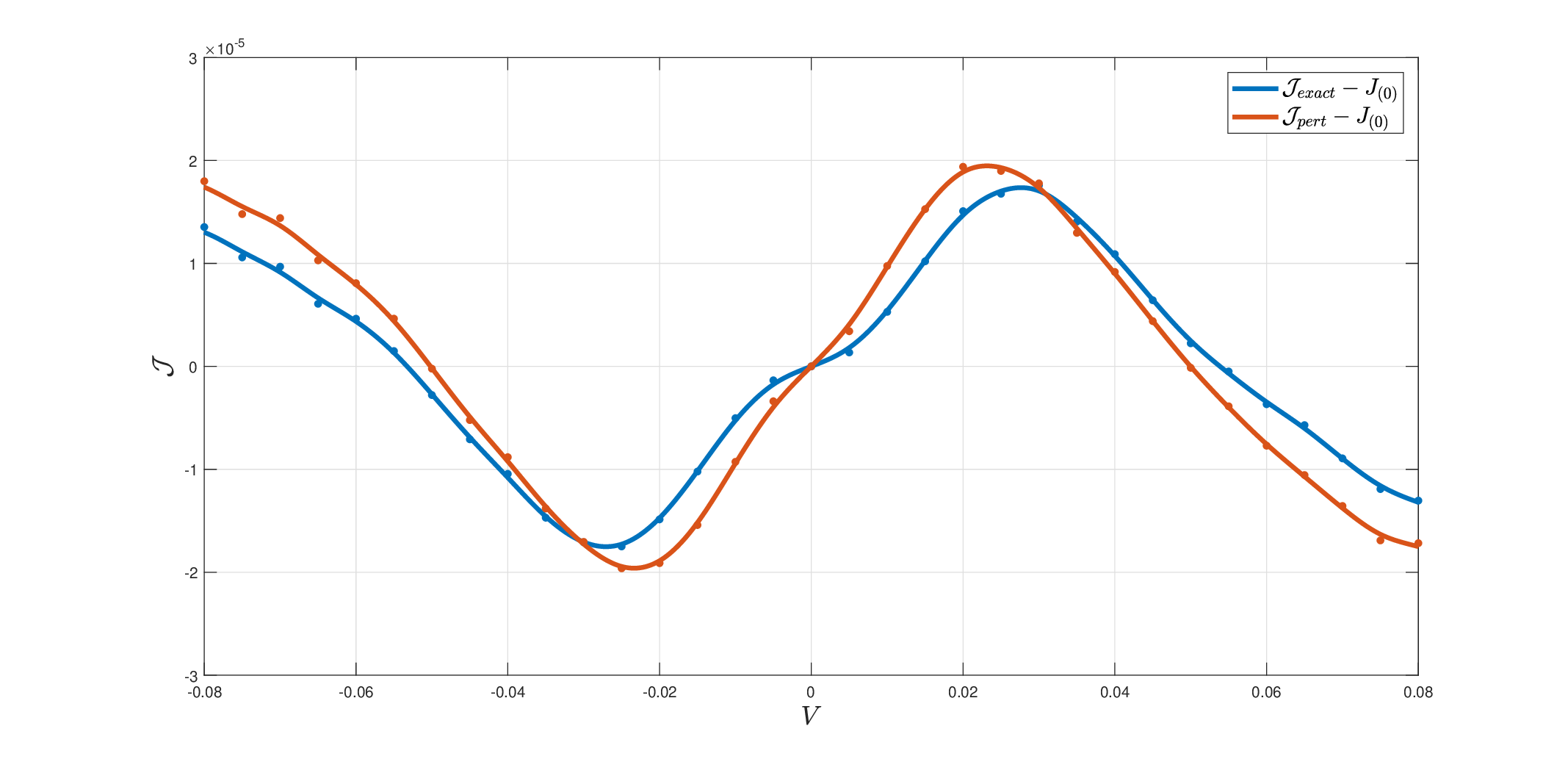}}}%
    \qquad
    \subfloat[]{{\includegraphics[trim=95 0 95 0, clip, width=8.5cm]{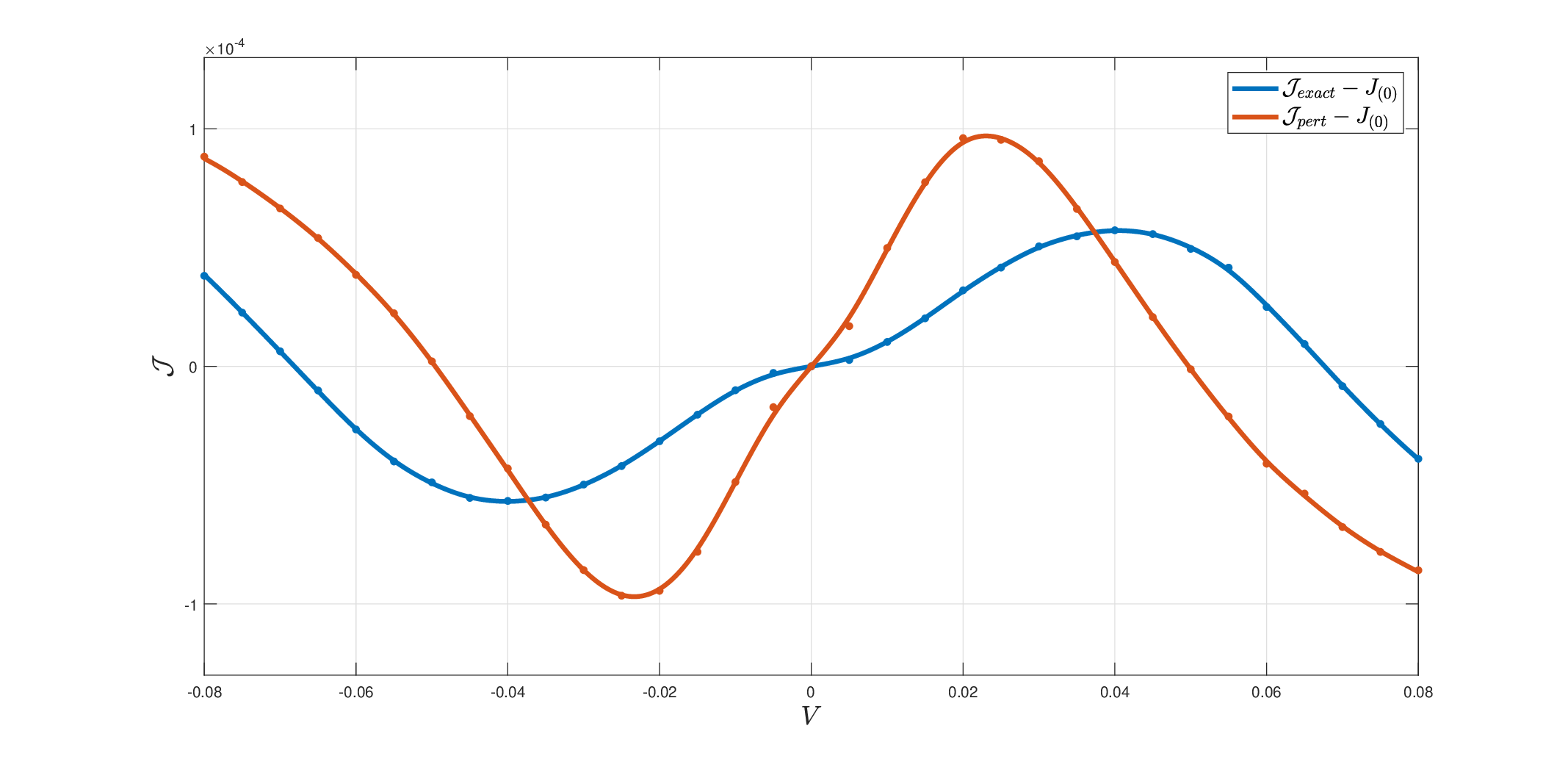}}}%
    \caption{Comparison of $\mathcal{J}_{\text{pert}} - J_{(0)}$ and $\mathcal{J}_{\text{exact}} - J_{(0)}$ as functions of voltage. Simulation is performed for the electronic parameters of $\Gamma_{L} = \Gamma_{R} = 0.01$, $\lambda = 0.05$, $k = 0.1$ and $V = 0.1$. In (a) we have the nuclear parameters of $m = 5000$ and (b) considers $m = 1000$. This gives a characteristic frequency of $\Omega = 0.0045$ for a) and $\Omega = 0.01$ for b).}
    \label{CorrectionsComparison}%
\end{figure*}

We assess the performance of our theory against the exact results by benchmarking the electronic current expressions in two ways: firstly, we calculate $\mathcal{J}_{\text{exact}}$ and $\mathcal{J}_{\text{pert}}$ as functions of time and compare these trajectories (see figure \ref{TrajectoriesComparison}); and secondly, we compute the current-voltage characteristics using the expressions for $\mathcal{J}_{\text{exact}}$ and $\mathcal{J}_{\text{pert}}$ (see figure \ref{CorrectionsComparison}). In the discussions that follow, a case of good quantitative agreement between the $\mathcal{J}_{\text{exact}}$ and $\mathcal{J}_{\text{pert}}$ trajectories and a case of bad quantitative agreement will be considered. The reader can refer to Appendix \ref{Numerical Algorithm Description} for a general description of the algorithm used. Throughout this section, and the sections to follow, we will leverage a common set of parameters that we have chosen to reflect typical values observed in molecular electronic junctions.

\

In figure \ref{TrajectoriesComparison}, we plot $\mathcal{J}_{\text{exact}}$ and $\mathcal{J}_{\text{pert}}$ as functions of time for a single Langevin trajectory once the system has reached the steady state in the case of good and bad qualitative agreement; we also plot the adiabatic current $J_{(0)}$ as calculated according to equations (\ref{landauer}) and (\ref{T0}). Both calculations use the same set of parameters but with different masses for the classical degree of freedom: $m = 5000$ to show the case of good agreement, and $m = 1000$ for the case of bad agreement.

\

Figure \ref{TrajectoriesComparison} a) shows that in the case of good qualitative agreement there is a perturbative current $\mathcal{J}_{\text{pert}}$ that approximates closely the exact current $\mathcal{J}_{\text{exact}}$; the exact current has been shifted slightly relative to the adiabatic current $J_{(0)}$ and this has been captured by the perturbative current. The case of bad qualitative agreement, on the other hand, shows that $\mathcal{J}_{\text{pert}}$ poorly approximates $\mathcal{J}_{\text{exact}}$, as evidenced by the $\mathcal{J}_{\text{pert}}$ trajectory drastically overshooting and undershooting the analytical $\mathcal{J}_{\text{exact}}$. For the positive voltages considered, $\mathcal{J}_{\text{pert}}$ and $\mathcal{J}_{\text{exact}}$ have time-averaged values that are shifted upwards slightly relative to that of the adiabatic trajectory, the $\mathcal{J}_{\text{pert}}$ curve having the highest time-averaged value. The most noticeable feature of this plot is that $\mathcal{J}_{\text{pert}}$ has a very large amplitude of oscillation relative to the other trajectories. 

\

Figure \ref{CorrectionsComparison} shows a plot of the current-voltage characteristics of the device in the case of good qualitative agreement and bad qualitative agreement. Simulations are performed for the same parameters as figure \ref{TrajectoriesComparison} by considering the motion corrections to the adiabatic current $J_{(0)}$ as calculated according to the perturbative current $\mathcal{J}_{\text{pert}}$ (as given by $\mathcal{J}_{\text{pert}} - J_{(0)}$) and the analytical current $\mathcal{J}_{\text{exact}}$ (as given by $\mathcal{J}_{\text{exact}} - J_{(0)}$). The current-voltage plot is calculated in a similar fashion to the trajectories in figure \ref{TrajectoriesComparison}; however, we compute these trajectories for an array of voltages and compute the mean and time-averaged currents once the system has thermalised. 

\

As demonstrated by figure \ref{CorrectionsComparison}, the case of good qualitative agreement exhibits much closer agreement between the time-averaged perturbative and analytical current-voltage characteristics relative to the case of bad qualitative agreement. The time-averaged perturbative current is noticeably shifted away from the time-averaged analytical current, with this being particularly manifest in the case of bad qualitative agreement.

\

The characteristic frequency for nuclear motion can be computed from the spring constant and mass as $\Omega=\sqrt{k/m}$, with $\Omega=0.0045$ for the case of good agreement and $\Omega=0.01$ for the case of bad agreement. Therefore, we see that, as qualitatively expected, if nuclear motion is slow compared to the electron tunnelling dynamics $\Omega \ll \Gamma$, the perturbative approach produces the almost exact results, however, when the nuclear electronic time-scales become comparable $\omega\sim \Gamma$ the perturbative current  deviates both qualitatively and quantitatively from exact values.

\subsection{Non-Adiabatic Corrections}
\label{Dynamical Blockade}
From our previous works we found that the presence of second-order corrections in the central-time derivatives produced a current-blockade effect in the current-voltage characteristics. Our work assumed that nuclear dynamics could be described in the equilibrium limit, meaning that we computed the average current according to the calculation
\begin{equation}
\big\langle \mathcal{J} \big\rangle = \int dx dp P(x, p) \mathcal{J}(x, p), 
\label{Boltzmann}
\end{equation}
where $\mathcal{J}(x, p)$ is the non-adiabatically corrected current computed for a given point in the phase space and
\begin{equation}
P(x,p) \sim e^{-p^2/2mT} e^{-U(x)/T}
\end{equation}
is the equilibrium Boltzmann distribution function. The discussion that follows will now extend this analysis to consider classical particles that can undergo non-equilibrium motion in accordance with the Langevin equation. Our discussion will start by considering the non-equilibrium probability distributions $P_{neq}(x)$ that result from non-equilibrium motion and, secondly, we will analyse the resultant non-equilibrium conductance-voltage characteristics.

\

For this set of electronic and nuclear parameters, the conductance-voltage characteristics are calculated in the following fashion (the reader is invited to view Appendix \ref{Numerical Algorithm Description} for a more thorough explanation): (i) for a given voltage $V$, $n$ stochastic nuclear trajectories are calculated according to the Langevin equation; (ii) from these trajectories the analytical and perturbative current trajectories are calculated; (iii) once a steady-state has been reached, the remaining current trajectories are averaged; (iv) steps (i) to (iii) are repeated for values across the voltage spectrum to produce the current-voltage profile.

\

\begin{figure*}
    \centering 
    \subfloat[]{{\includegraphics[width=0.3\textwidth]{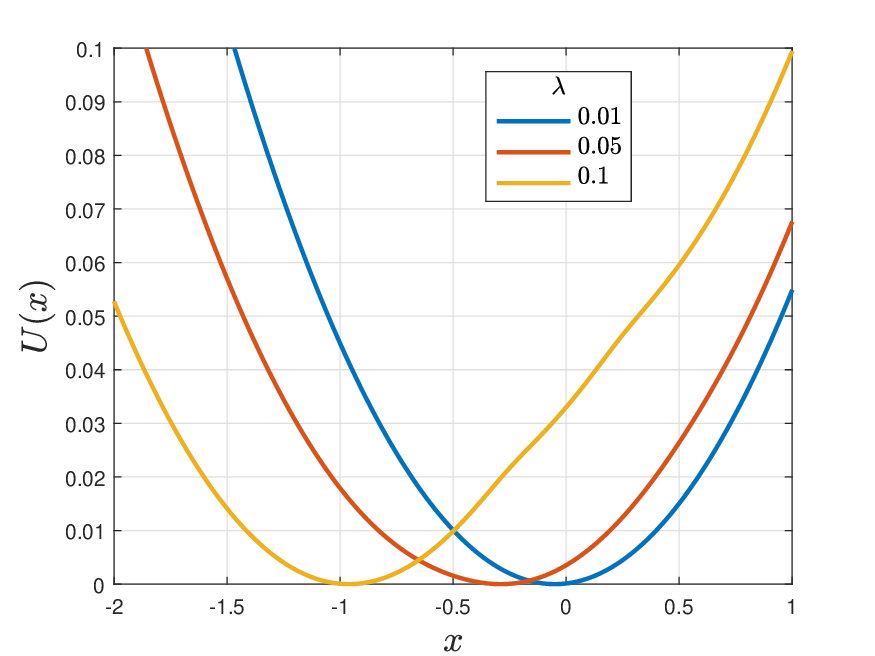}}}%
    \qquad
    \subfloat[]{{\includegraphics[width=0.3\textwidth]{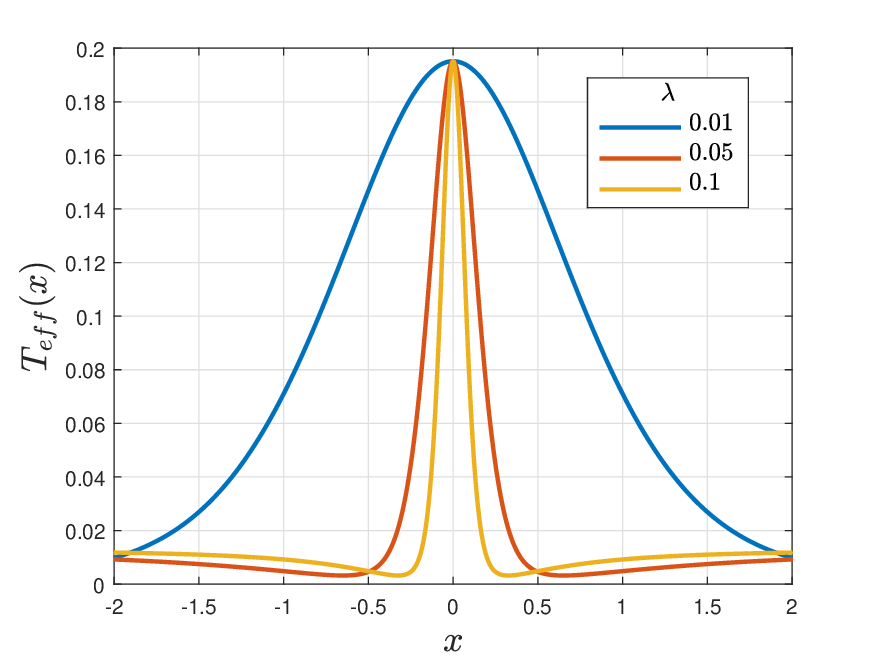}}}%
    \qquad
    \subfloat[]{{\includegraphics[width=0.3\textwidth]{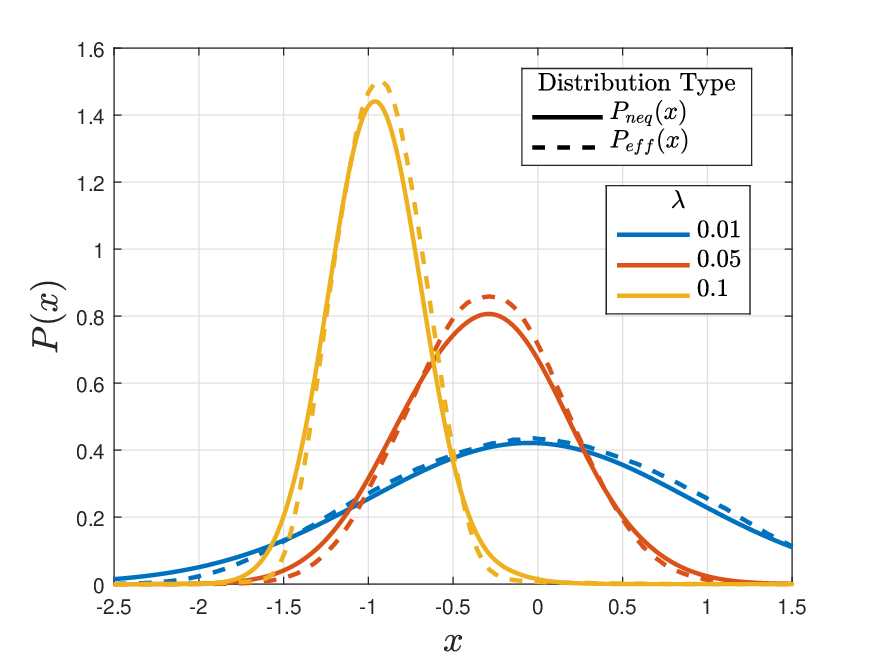}}}%
    \caption{(a) Adiabatic effective potentials $U_{eff}(x)$, (b) effective temperatures $T_{eff}(x)$ and (c) a comparison of the probability distributions $P_{neq}(x)$ and $P_{eff}(x)$. All calculations use the parameter set $k = 0.1$, $\Gamma_{L} = \Gamma_{R} = 0.01$, $h_{0} = 0$, $V = 0.05$, $m = 5000$ and values $\lambda = 0.01$, $\lambda = 0.05$ and $\lambda = 0.1$. The spring-constant $k$ and the mass $m$ give a characteristic frequency of $\Omega = 0.0045$.}
    \label{PotentialsandTemps}
\end{figure*}

It will be useful in what follows to use an effective temperature quantity $T_{eff}(x)$:
\begin{equation}
T_{eff}(x) = \frac{D(x)}{2 \xi(x)},
\label{Teff}
\end{equation}
where $D(x)$ and $\xi(x)$ are the previously-defined diffusion and friction coefficients. The introduction of local effective temperature quantities is not a universally valid approach and has been known to fail in many nanoscopic systems \cite{Efimkin2018}. Nonetheless, we find that the effective temperature serves as a useful ratio to infer the relative strengths of the random forces and frictional forces for a given nuclear displacement $x(t)$ and voltage $V$ and as a predictor of steady-state characteristics. Additionally, it will be useful to use an effective adiabatic potential. This quantity describes the 'effective' classical potential that arises from the originally defined classical potential $U(x)$ and the adiabatic force $F(x)$ in the Langevin equation:
\begin{equation}
U_{eff}(x) = U(x) - \int^{x}_{0} d x^{\prime} F(x^{\prime}).
\label{ueff}
\end{equation}
The reader can refer to figure \ref{PotentialsandTemps} a) and b) for a visual representation of $U_{eff}(x)$ and $T_{eff}(x)$ for the parameters selected. Considering figure \ref{adiabaticforces} a), we see that the effective potential tends to the external pinning potential $U(x)$ in the case of lower $\lambda$ and is deformed to the LHS with increasing $\lambda$. The deformation is a result of $\lambda$ being positive and the energy level $h$ lowering with decreasing $x$ (see equation \ref{Hami}) and, via equation \ref{adiabaticforces}, the adiabatic force increases due to the energy-level occupation term inside it. This makes sense as altering $\lambda$ will weaken/strengthen the coupling between the nuclear and electronic systems. Figure \ref{adiabaticforces} b), however, shows effective temperature profiles that become more pronounced for higher $\lambda$, all with a central peak at $x = 0$. The peak is a result of the molecular orbital being in a state of resonance at $x = 0$. The broadening of the effective temperature profile is a result of $h$ being less dependent on $x$; as we slide $x$ from $-2$ to $2$, the energy level $h$ will more rapidly slide across the energy space with increase $\lambda$.

\

Figure \ref{PotentialsandTemps} c) considers two sets of probability density functions for systems with electron-vibrational  couplings $\lambda$ of varying strength. The first set, given by $P_{neq}(x) $ (the solid curves), are the non-equilibrium probability density functions; and the second set are what we will refer to as the effective probability density function $P_{eff}(x)$ (the dashed curves). The density functions $P_{neq}(x)$ were generated by running many Langevin trajectories and building a normalised frequency distribution $P_{neq}(x)$ for each value of $\lambda$. The effective probability density function $P_{eff}(x)$, on the other hand, is defined according to the formula:
\begin{equation}
P_{eff}(x) = \frac{1}{Z} e^{-\frac{U_{eff}}{\langle T_{eff} \rangle}}.
\end{equation}
Above there is the adiabatic potential $U_{eff}$ and the steady-state effective temperature $\langle T_{eff} \rangle$ relevant to each value of $\lambda$. Additionally, above we have the appropriate normalisation constant $Z$. The steady-state effective temperature $\langle T_{eff} \rangle$ is computed by running long trajectories and computing an average over time once a steady state has been reached; in addition, many long trajectories are run and these results are further averaged. 

\

As can be seen in Figure \ref{PotentialsandTemps} c), the non-equilibrium distributions have non-zero expectation values $\langle x \rangle < 0$ that move further to the left with increasing $\lambda$. In addition, we see that the variance of these probability distributions decreases with increasing $\lambda$. This phenomenon can be explained in large part by the adiabatic force $F$ and its increasing prevalence with increasing $\lambda$. For the parameter set selected, the adiabatic force tilts the effective adiabatic potential to the left and, consequently, influences the classical degree of freedom to spend most of its time there. In addition, the effective temperature $T_{eff}(x)$ is at a maximum at $x = 0$ and the particle will be influenced to move into the colder regions that are now provided by the changing effective potential $U_{eff}(x)$. Another interesting point to note from Figure \ref{PotentialsandTemps} c) is the close agreement between $P_{neq}(x)$ and $P_{eff}(x)$. This agreement has been found for a variety of parameter sets and suggests that the effective temperature $T_{eff}(x)$ can be used as an effective measure of steady-state temperature characteristics. 

\

We now turn our attention to analysing the conductance-voltage characteristics and contrasting the differences that arise in the equilibrium and non-equilibrium descriptions. As already mentioned, our previous work made use of an equilibrium description of the dynamics in accordance with a Boltzmann factor \cite{Kershaw2017a,Kershaw2018a,Kershaw2019}. In this equilibrium limit, it can be shown that the transmission (\ref{simplifiedSAPC}) reduces to
\begin{equation}
\big\langle \mathcal{T}_{L} \big\rangle = - \frac{1}{\pi} \frac{ \Gamma_{L} \Gamma_{R}}{\Gamma_{L} + \Gamma_{R}} \text{Im} \Big\{ G^R - \frac{3 k T}{m} \lambda^2 \big( G^{R} \big)^5 \Big\},
\label{simplifiedSAPC}
\end{equation}
where we emphasise that the brackets indicate averaging over nuclear velocities in accordance with the Boltzmann factor. An important consequence of this description is that terms linear in nuclear velocities and accelerations disappear once averaged and only corrections that are quadratic in nuclear velocities will impact the conductance-voltage characteristics.

\begin{figure*}
    \centering 
    \subfloat[]{{\includegraphics[trim=0 0 0 20, clip, width=0.3\textwidth]{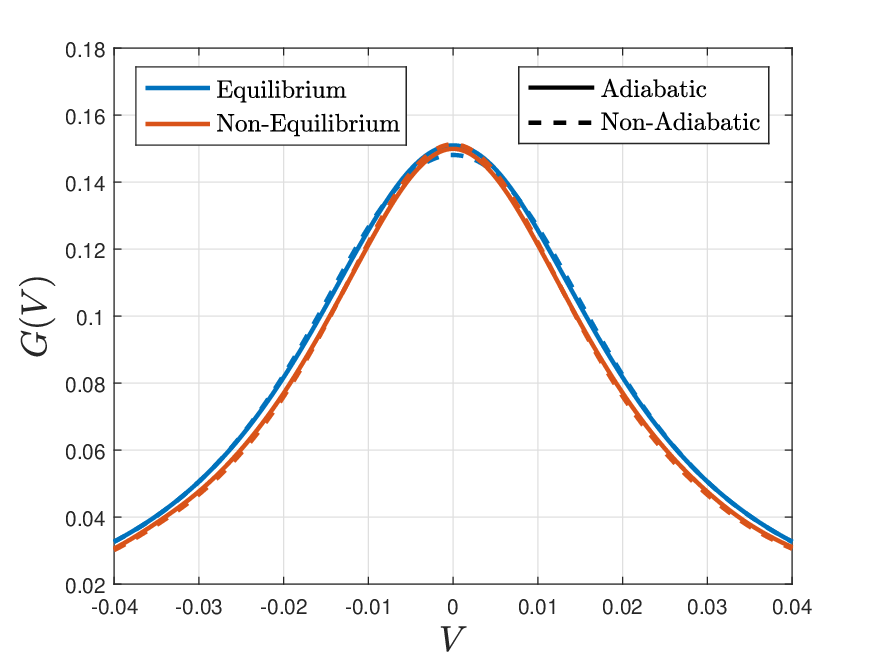}}}
    \qquad
    \subfloat[]{{\includegraphics[trim=0 0 0 20, clip, width=0.3\textwidth]{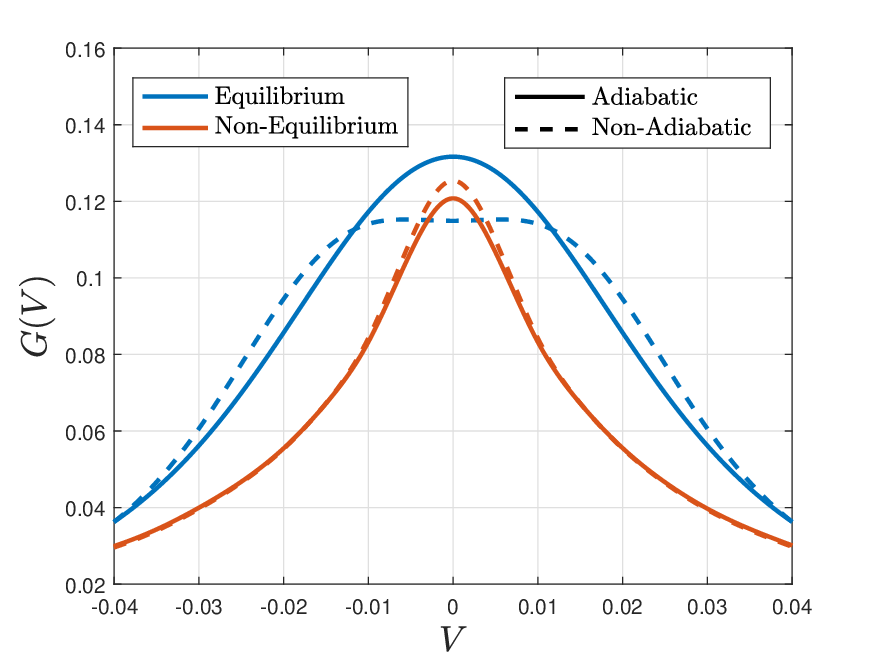}}}
    \qquad
    \subfloat[]{{\includegraphics[trim=0 0 0 20, clip, width=0.3\textwidth]{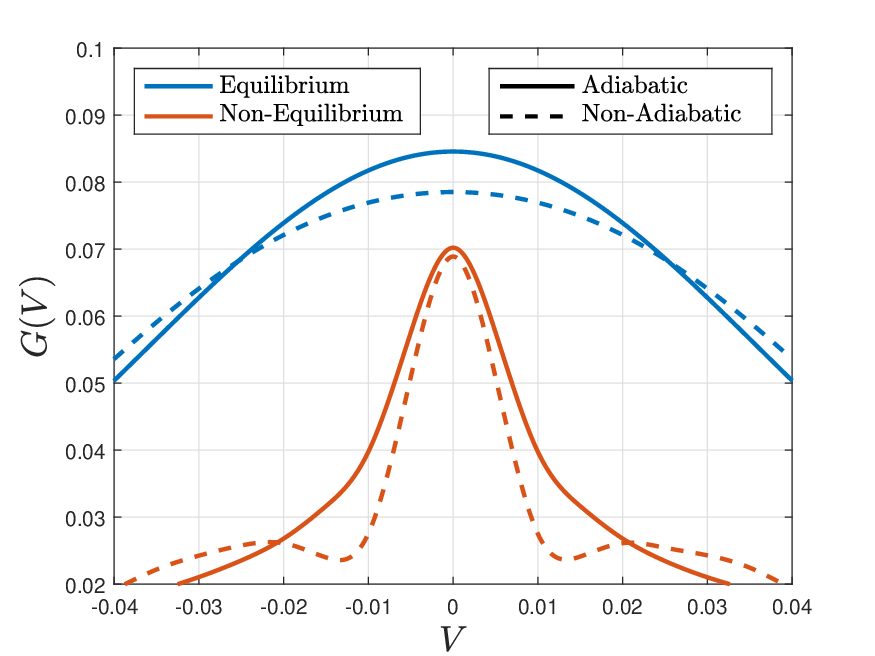}}}
    \caption{Adiabatic conductance (blue curve) and non-adiabatic conductance (orange curve) profiles for the Boltzmann description. All plots share the parameter set $\Gamma_{L} = \Gamma_{R} = 0.01$, $k = 0.1$ and $m = 5000$; however, plots a), b) and c) differ through $\lambda$ values of $\lambda = 0.01$, $\lambda = 0.05$ and $\lambda = 0.1$, respectively. The spring-constant $k$ and the mass $m$ give a characteristic frequency of $\Omega = 0.0045$.}
    \label{BoltzmannPlots} 
\end{figure*}

\

Figure \ref{BoltzmannPlots} has been generated to enunciate the differences between equilibrium and non-equilibrium motion. The plots consider the equilibrium case as the blue curve and non-equilibrium motion as the orange curve; in addition, the adiabatic and non-adiabatic conductances are given by the solid and dashed lines, respectively. The first thing to notice is the difference in the adiabatic conductances between the two descriptions. These plots show in all cases that the adiabatic conductance as calculated according to the Langevin equation is less than the adiabatic conductance for the equilibrium description and, furthermore, that the discrepancy between the two increases with increasing $\lambda$. We have found that this difference is due to the position component of the non-equilibrium probability density distribution $P_{neq}(x)$ (as generated according to the non-equilibrium description) being broader than the Boltzmann component $P_{eq}(x) = e^{\frac{-U(x)}{kT}}$; additionally, we have found that the discrepancy between $P_{neq}(x)$ and $P_{eq}(x)$ increases with increasing $\lambda$. Thus, the conclusion is that the broader non-equilibrium probability density function $P_{neq}(x)$ in conjunction with the current surface $J_{(0)}$ produces a conductance profile that is lower than the equilibrium profile.

\

When analysing the non-adiabatic conductance of figure \ref{BoltzmannPlots}, we see that non-equilibrium  nuclear dynamics have a significant effect on molecular junction electron transport properties. First, the non-equilibrium nuclear motion reduces the differential conductance relative to the equilibrium case. These effects are exacerbated by changing the values of $\lambda$. Second, the second-order non-adiabatic corrections act differently in equilibrium and non-equilibrium by increasing or reducing the 
electronic conductivity. The opposite role for non-adiabatic corrections for non-equilibrium and equilibrium nuclear dynamics can be attributed to the acceleration dependent term in (\ref{transmission_corrections}) which becomes particularly important for strongly coupled non-equilibrium electron-nuclear dynamics.

\subsection{Landauer Blowtorch Effect and Current-Induced Bi-Stability}
\label{Blowtorch Effect}
\begin{figure*}
    \centering 
    \subfloat[]{{\includegraphics[width=0.3\textwidth]{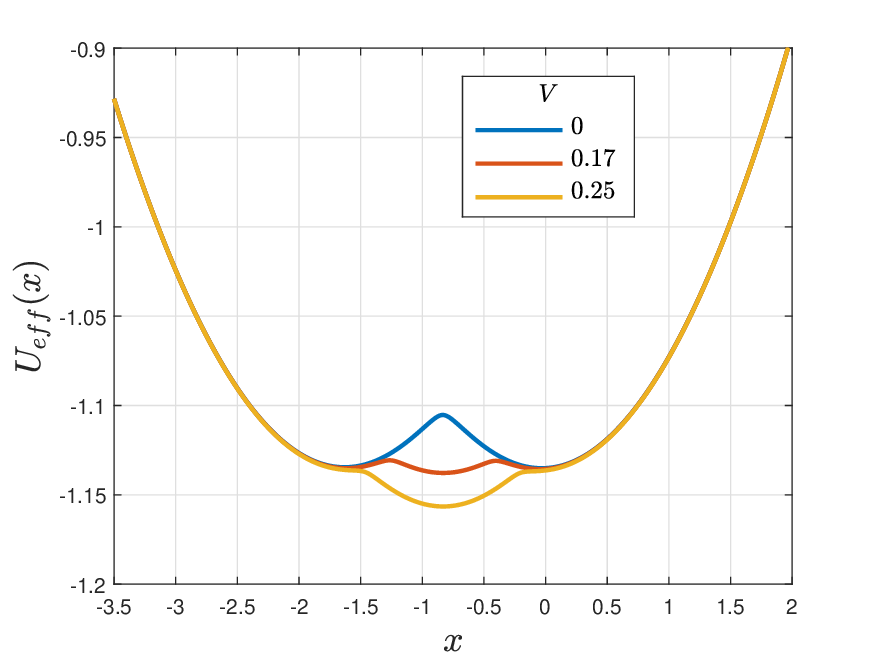}}}%
    \qquad
    \subfloat[]{{\includegraphics[width=0.3\textwidth]{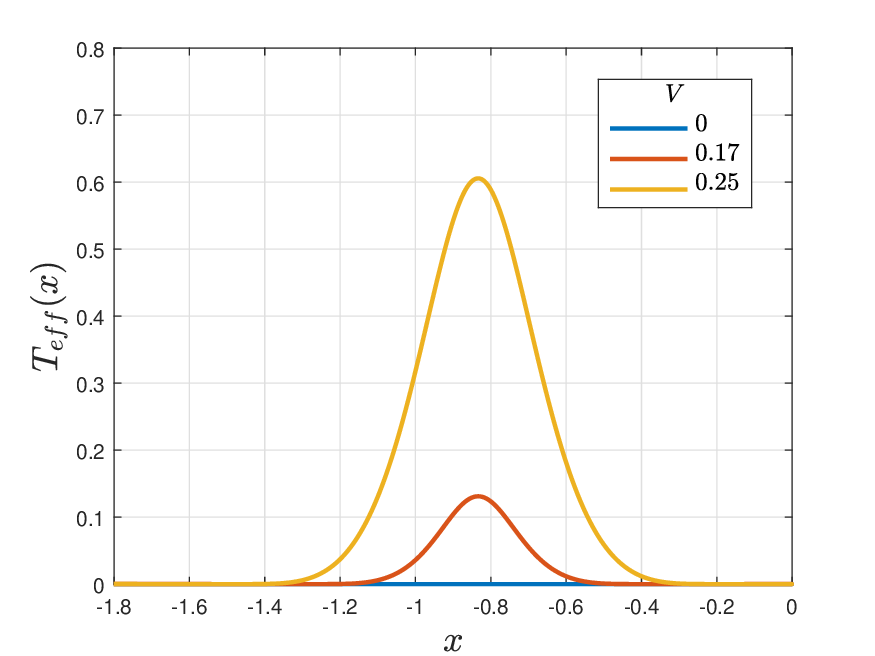}}}%
    \qquad
    \subfloat[]{{\includegraphics[width=0.3\textwidth]{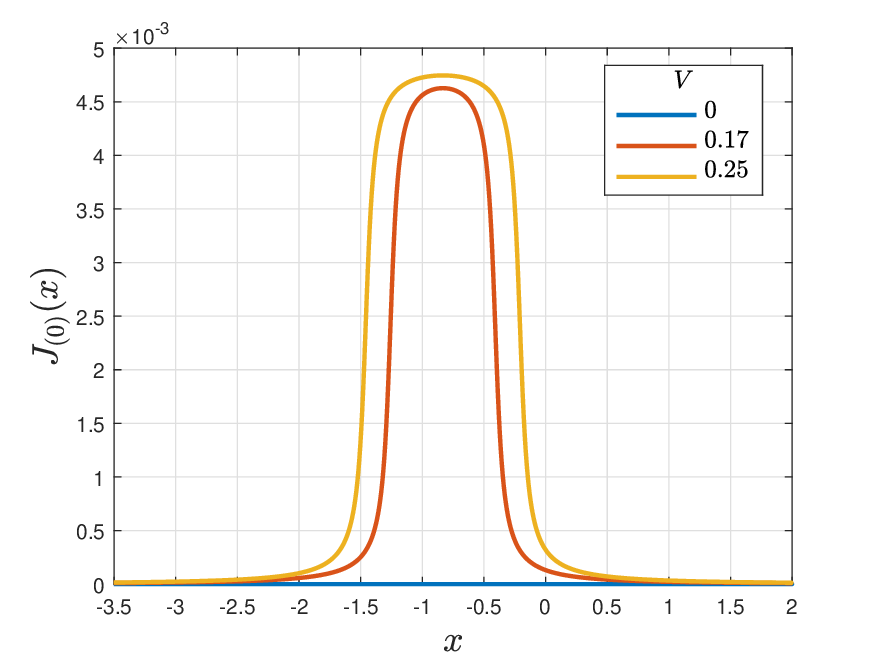}}}%
    \caption{(a) Adiabatic effective potentials $U_{eff}(x)$, (b) effective temperatures $T_{eff}(x)$ and (c) a comparison of the adiabatic current profiles $J_{(0)}(x)$. All calculations use the parameter set of a spring constant of $k = 0.06$ and the electronic parameters of $\Gamma_{L} = \Gamma_{R} = 0.01$, $h_{0} = 0.17$ and $\lambda = 0.2$. Each plot has three curves that correspond to the voltages $V = 0$, $V = 0.17$ and $V = 0.25$. A mass of $m = 5000$ was used to give a characteristic frequency of $\Omega = 0.0035$.}
    \label{PotentialTemperatureDistribution}
\end{figure*}

This section discusses an interesting phenomenon that arises in the non-equilibrium Langevin description: the natural appearance of the blowtorch effect and an important role it plays in current-induced switching between molecular configurations. The blowtorch effect is the kinetically controlled dynamics via elevation of the temperature for a portion of the potential energy surface \cite{LANDAUER1993551,PhysRevA.12.636}.
As we shall now demonstrate, the Langevin description of nuclear dynamics results in steady-state nuclear temperatures that are non-constant with nuclear displacement $x(t)$ and voltage $V$. These variations in steady-state temperature with displacement and voltage allow the possibility, among other things, of designing devices with voltage-dependent thermal characteristics and thermally-activated switching. 
We once again find it useful for the purposes of analysis and discussion to work with the effective temperature quantity $T_{eff}(x)$ defined via (\ref{Teff}). We have provided \ref{PotentialTemperatureDistribution} b) to see a visualisation of the effective temperature surface as a function of nuclear displacement $x$ and voltage $V$. In addition, the adiabatic current $J_{(0)}$ has been plotted as a function of position and voltage in figure \ref{PotentialTemperatureDistribution} c): this plot will prove to be useful when interpreting the Fano factor. 

\

This section will also make use of the effective adiabatic potential (\ref{ueff}). Figure \ref{PotentialTemperatureDistribution} a) shows for the parameters selected that increasing the voltage has the effect of changing the number of minima that the potential has. For a voltage of $V = 0$, the effective potential takes the form of a double potential as seen in figure \ref{PotentialTemperatureDistribution} a). The second potential arises via a combination of the original pinning potential $U(x)$ (the RHS potential) and the adiabatic force $F$. Our choice of positive $\lambda$ and non-zero $h_{0}$ means that an additional potential forms in the negative regions of the $x$-axis (a positive non-zero value for $h_{0}$ will mean that the adiabatic force occurs for negative $x$-values as the energy level $h$ is shifted relative to $x = 0$). Increasing the voltage results in the formation of a central minima and the effective potential turns into a triple potential, an example of which is provided in figure \ref{PotentialTemperatureDistribution} a) for a voltage of $V = 0.17$. Increasing the voltage further sees the effective potential morph into a single potential as the central potential dominates. Figure \ref{PotentialTemperatureDistribution} b) shows, for the parameters selected, the relative effective temperatures as a function of position. By comparing figure \ref{PotentialTemperatureDistribution} a) and b), we see a uniform temperature profile across the potential (the temperature is the background bath temperature in accordance with the fluctuation-dissipation theorem) where, upon increasing the voltage, we see that the central minima achieves relatively higher effective temperatures compared to its neighbouring wells.  

\

\begin{figure*}
    \centering
    \subfloat[]
 {{\includegraphics[width=0.3\textwidth]{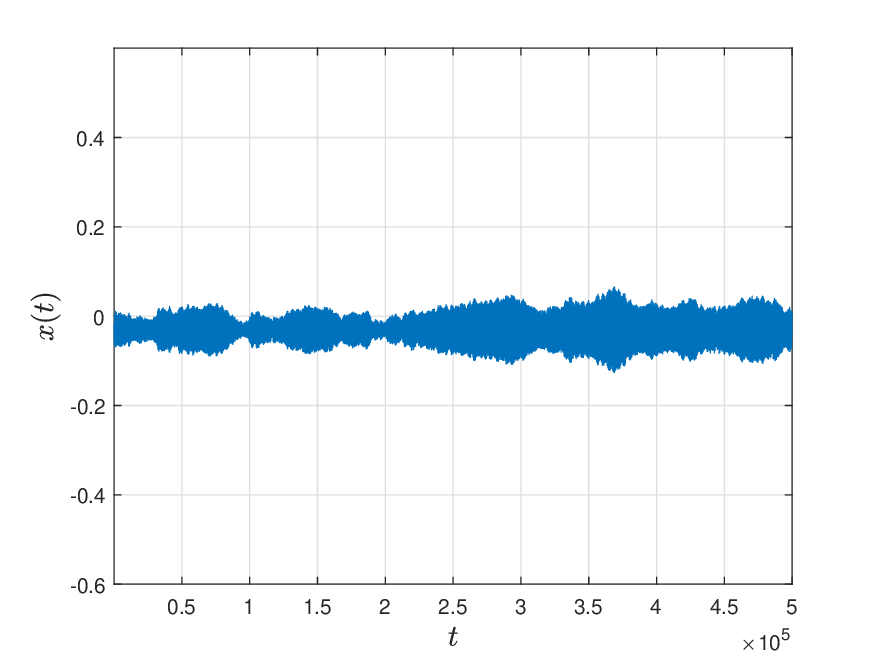}}}
    \qquad
    \subfloat[]
{{\includegraphics[width=0.3\textwidth]{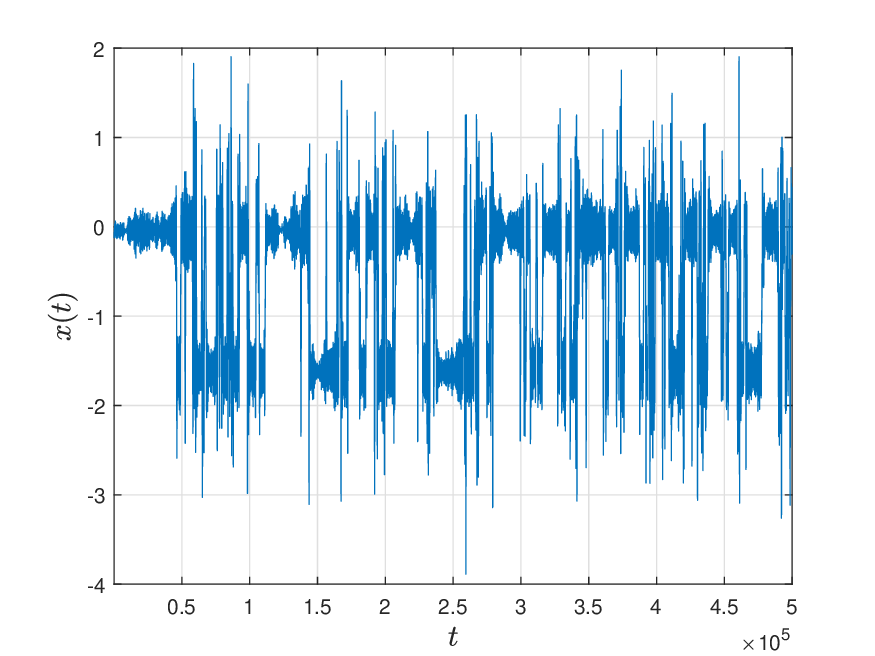}}}
    \qquad
    \subfloat[]
{{\includegraphics[width=0.3\textwidth]{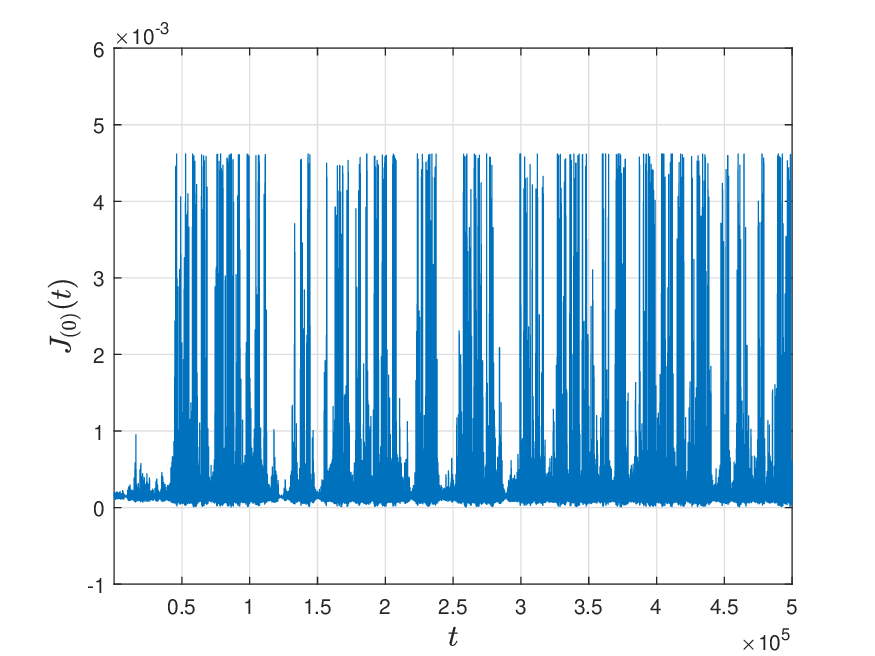}}}
    \caption{(a) Trajectory of the classical coordinate $x(t)$ as a function of time $t$ for a voltage of $V = 0$, (b) trajectory of the classical coordinate $x(t)$ as a function of time $t$ for a voltage of $V = 0.25$ and (c) the current $J_{(0)}(t)$ as a function of time $t$ for a voltage of $V = 0.25$. Simulation is performed for the parameters of $k = 0.06$, $\Gamma_{L} = \Gamma_{R} = 0.01$, $h_{0} = 0.17$ and $\lambda = 0.2$. A mass of $m = 5000$ was used to give a characteristic frequency of $\Omega = 0.0035$.}%
    \label{SwitchingTrajectory}%
\end{figure*}

We now analyse the motion of a classical degree of freedom $x(t)$ according to a Langevin description. Figure \ref{SwitchingTrajectory} a) and b) show a simple Langevin trajectory $x(t)$ as a function of time: in a) the simulation is performed in the limit of $V = 0$ and b) considers a trajectory for $V = 0.17$. Interpreting this in terms of our effective potentials and our effective temperatures, figure \ref{SwitchingTrajectory} a) corresponds to a double potential with a uniform temperature profile as seen in figure \ref{PotentialTemperatureDistribution} a) and b) for $V = 0$ and figure \ref{SwitchingTrajectory} b) corresponds to those profiles of voltage $V = 0.17$. 

\

Figure \ref{SwitchingTrajectory} a) shows a nuclear trajectory that is constrained to oscillate within the previously-described system for a voltage of $V = 0$ and b) shows a voltage of $V = 0.17$. We see that the classical particle is restrained to oscillate within one of the local minima in $U_{eff}(x)$ without an applied voltage: the potential barrier between the two wells and the low local temperatures does not provide enough energy to the particle to make a switch. Non-zero voltages, on the other hand, show a particle that mimics switching behaviour between the far-left well and the far-right well. The potential barriers between wells and the local temperature provide enough energy for a particle to make a transition; note, however, how the particle mimics the switching behaviour of a double potential well as it predominantly switches between the far-left and the far-right wells. This is due to the high temperatures in the central potential relative to the low temperatures in the neighbouring wells: a particle entering the central potential will gather enough energy to make a transition much more quickly than a particle entering the far-left or the far-right wells. The corresponding current trajectory in \ref{SwitchingTrajectory} c) shows non-zero current for a voltage of $V = 0.17$ and 'spikes' of non-zero current. This can be explained by observing the effective adiabatic potentials in figure \ref{PotentialTemperatureDistribution}, where we see that the left and right minima correspond to regions of zero current for the parameters selected. On the other hand, nuclear positions of around $x = - 1$ correspond to regions of non-zero current. A classical particle trapped in the left or right potential will therefore result in zero current flowing across the system and it is the switching of the classical particle among the left and right well, and its localisation to the central well, that will produce non-zero current. Notice that this 'current window' widens somewhat linearly with voltage; furthermore, notice how the magnitude of the current increases with the voltage: for low voltages the magnitude of the current is relatively sensitive to the voltage in comparison to high voltages. In summary, switching of the classical particle amongst the wells corresponds to the switching of the electric current on and off: increasing the voltage will increase the current window (the portion of the trajectory that corresponds to non-zero current) and the magnitude of the current, and the formation of the central well will further increase the portion of the trajectory that corresponds to non-zero current.  

\

\begin{figure*}
    \centering 
    \subfloat[]
    {{\includegraphics[width=0.45\textwidth]{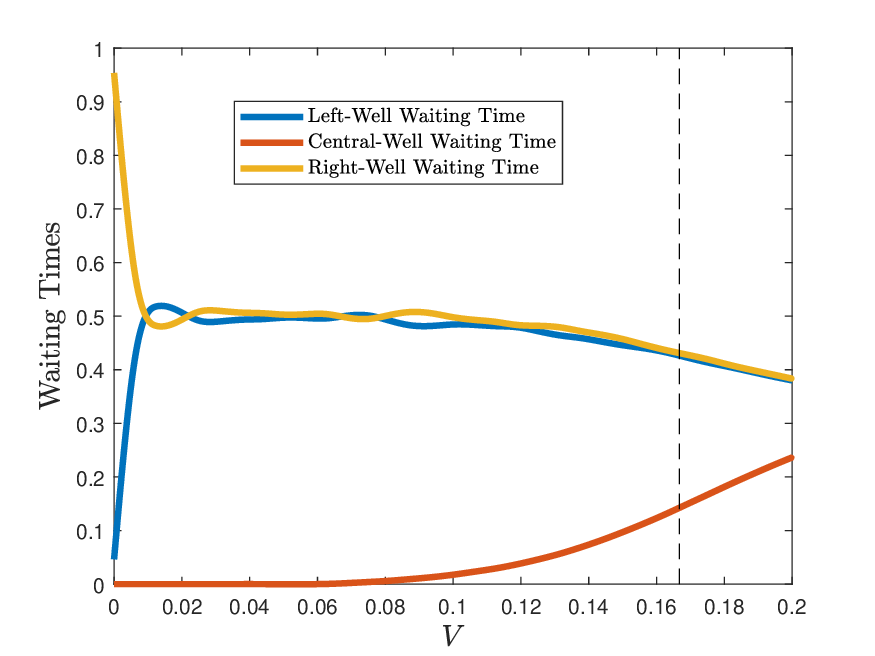} }}%
    \qquad
    \subfloat[]
    {{\includegraphics[width=0.45\textwidth]{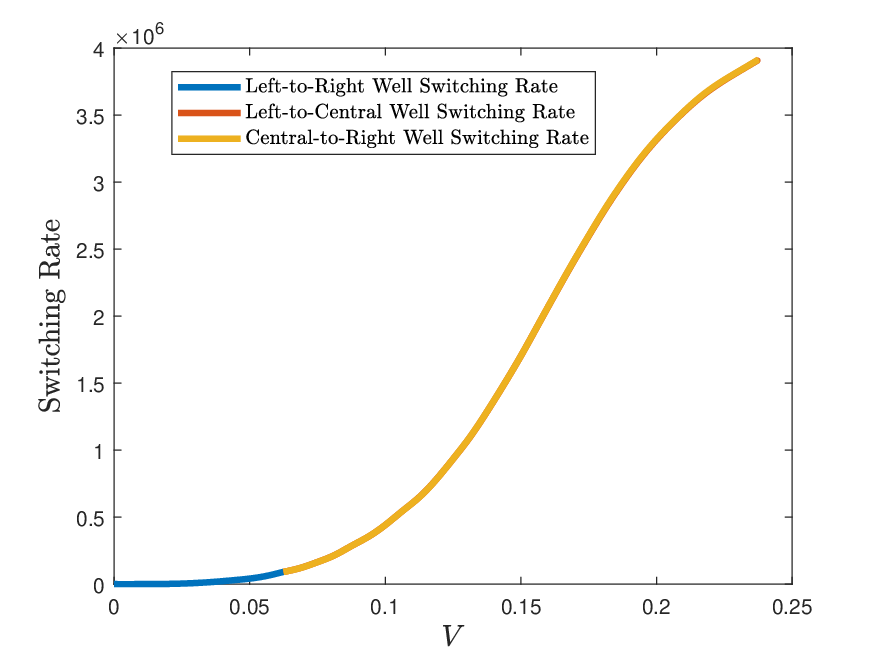} }}%
    \caption{In a) we have the waiting times of the classical particle in the far-left well, the central well and the far-right well where waiting times are presented in terms of a ratio from $0$ to $1$. In b) there is the switching rates of the classical particle amongst the three wells.}%
    \label{WaitingTimesTriple}%
\end{figure*}

The switching behaviour of the classical degree of freedom $x(t)$ amongst the wells as a function of voltage is more succinctly expressed in terms of the waiting times between the three wells, as seen in figure \ref{WaitingTimesTriple}. We define the waiting times as the time spent in each of the respective wells: for example, if the left and right wells have waiting times of 0.2 and 0.8 respectively, then this would indicate that, for a given trajectory, the particle spent 20\% of its time in the left well and 80\% of its time in the right well. Calculations of the waiting times separate the wells for a given effective potential according to the maxima that occur. 

\

Voltages ranging from $V = 0$ to approximately $V = 0.06$ correspond to the effective adiabatic potential $U_{eff}(x)$ emulating a double potential; voltages in the range of $V = 0$ to approximately $V = 0.01$ result in the nuclei remaining localised in a single well (the single well that the particle remains in corresponds to the initial condition set for the particle), while voltages ranging from $V = 0.01$ onwards demonstrate the particle accessing multiple wells. As we increase the voltage to a point greater than approximately $V = 0.06$, the effective adiabatic potential begins to look like a triple potential and the particle begins to spend an appreciable amount of time in the newly-formed central potential. The higher the voltage, the more prominent the central potential becomes, and we see the particle spends increasing amounts of time in the central potential. The horizontal line in figure \ref{WaitingTimesTriple} corresponds to a voltage of $V = 0.17$ and the effective adiabatic potential given by figure \ref{PotentialTemperatureDistribution}. This scenario corresponds to the triple potential having three minima of approximately equal height; note, however, that the particle spends minimal time in the central potential due to the high temperatures. We also note for increasing voltages that the time in the far-left and far-right wells is split evenly as a consequence of the symmetry of the well and its temperature profile.  

\

We finish section \ref{Blowtorch Effect} by considering current noise. We compute the current-current autocorrelations and the Fano factor as a function of voltage. We calculate $n$ Langevin trajectories for a given voltage.  For a given trajectory $i$, we compute  the deviation from the mean according to the formula: 
\begin{equation}
\delta \mathcal{J}_{i} (t) = \mathcal{J}_{i} (t) - \text{lim}_{T \rightarrow \infty} \frac{1}{T} \int^{T}_{0} dt \mathcal{J}_{i}(t).
\end{equation}
Note that we are using the variable $\mathcal{J}$ generally for any current quantity of interest: for the calculations that follow, we will take $\mathcal{J}$ to be the perturbative current $\mathcal{J}_{pert}$ and the adiabatic current $\mathcal{J}_{(0)}$. We then compute the  current noise due to nuclear motion according to the expression:
\begin{equation}
S_{\alpha} (\tau) = 2 \text{lim}_{T \rightarrow \infty} \frac{1}{T} \int^{T}_{0} d t \delta \mathcal{J}_{i} (t) \delta \mathcal{J}_{i} (t + \tau),
\end{equation}
where we have ignored quantum mechanical  cross terms in the calculation. This in turn allows us to calculate the Fano factor $F_{i}$ for the $i^{th}$ Langevin trajectory according to:
\begin{equation}
F_{i} = \frac{S_{i} (\omega = 0)}{2 \langle \mathcal{J}_{i} \rangle}.
\end{equation}
Above we have the time-averaged current $\langle \mathcal{J}_{i} \rangle$ and the Fourier transform of the mechanical noise $S_{i} (\omega = 0)$ as evaluated for $\omega = 0$. Depending on the requirements of the calculation, the aforementioned procedure can be replicated for all of the $n$ Langevin trajectories to produce an averaged Fano factor:
\begin{equation}
F = \frac{1}{n} \sum_{i}^{n} F_{i}.
\end{equation}
If the Fano factor is smaller than 1 ($F<1$), the electron transport is sub-Poissonian statistical process; if $F=1$, it is Poissonian; if $F>1$, the transport is super-Poissonian.

\

\begin{figure*}
    \centering
    \vspace*{-0.75cm}    
    \includegraphics[width=0.7\textwidth]{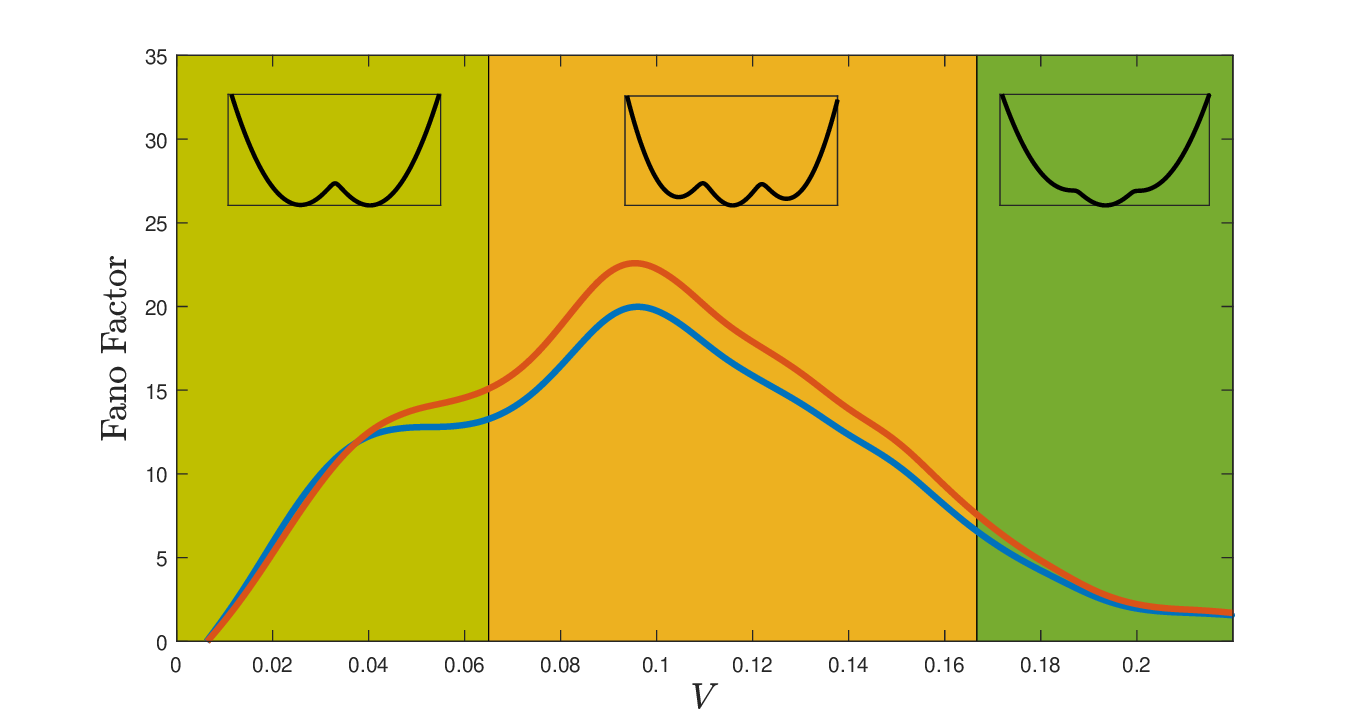}%
    \caption{A comparison of the adiabatic Fano factor and the non-adiabatic Fano factor as a function of voltage. Simulation is performed for electronic parameters of $\Gamma_{L} = \Gamma_{R} = 0.01$, $h_{0} = 0.17$ and $\lambda = 0.2$. Nuclear parameters are $k = 0.06$ and $m = 5000$.}%
    \label{FanoFactor}%
\end{figure*}

We  plot of the Fano factor as a function of voltage in figure \ref{FanoFactor} for the previously-described parameters. The plot consists of the adiabatic contribution to the Fano factor and the total Fano factor (as calculated by correcting the adiabatic Fano factor to the second-order in the central-time derivatives), as provided by the blue and orange curves, respectively. For the purposes of explanation the plot has been separated into several distinct regions, each colour coded and separated by a vertical line: the red region farthest to the left indicates the voltages for which the effective potential is a double potential; the orange region denotes voltages that generate triple potentials; and the blue region indicates when the system parameters have formed a single potential.  

\

Considering now the adiabatic Fano factor in figure \ref{FanoFactor}, we see that voltages in the range of $V = 0$ to approximately $V = 0.01$ are accompanied by Fano factors of approximately zero. The reason for this is intuitive: for the voltage range of $V = 0$ to approximately $V = 0.01$, the motion of the classical degree of freedom is localised to a single potential (see figure \ref{WaitingTimesTriple} for the waiting times) due to a shortage of energy that would be required for switching, leading to trajectories that are poorly autocorrelated. As the voltage is increased from $V = 0.01$ to around $V = 0.04$, we see that the Fano factor increases in a linear fashion. We explain this behaviour according to positive contributions of the autocorrelations outweighing the negative contributions in the following manner: as the voltage is increased from $V = 0.01$ to around $V = 0.04$, the current window is steadily increasing in width in a linear fashion with voltage and the current flowing across the junction is rapidly increasing with voltage, ultimately leading to rapidly increasing contributions; however, the switching rate is relatively low in this region (and is also seemingly increasing linearly with voltage) and negative contributions to the autocorrelations are consequently small.

\

Voltages of $V = 0.04$ to $V = 0.065$ display a Fano factor that remains relatively constant with voltage. This voltage region still sees the current window linearly increasing with voltage; however, the sensitivity of the current with voltage is now steadily decreasing and the switching rate is increasing more rapidly. The result is autocorrelations with increasingly dominant negative contributions that outweigh the positive contributions and flatten out the curve. Increasing the voltages further above $V = 0.065$ will see the double potential forming into a triple potential, the central potential forming within the current window (see a) of figure \ref{PotentialTemperatureDistribution}). The particle can now be trapped in the central potential and spends increasing amounts of time in the central potential with voltage (see figure \ref{WaitingTimesTriple}). The result is the junction producing non-zero current for increasing amounts of time and the positive contributions to the autocorrelations increase. The switching rate continues to increase; although, its negative contributions are outweighed by the previously-described positive contributions and the result is an increasing Fano factor to approximately $V = 0.1$. Increasing the voltages beyond $V = 0.1$ results in switching rates that begin to dominate the autocorrelations: the negative contributions continually overpower the positive contributions and the Fano factor drops. 

\newpage
\section{Conclusions}
In this paper we have extended our description of electronic transport in molecular junctions and its interaction with nuclear degrees of freedom. Using the NEGF formalism and a Wigner expansion to extract and separate fast and slow time-scales, we describe the electronic transport across the junction according to an extended current formula that accounts for the velocities, kinetic energies and accelerations of the classical degrees of freedom. In addition, we describe the nuclear degrees of freedom according to a set of Langevin equations with the adiabatic force, frictional coefficient and diffusion coefficients that are derived self-consistently within the NEGF formalism. 

\

As a first step in applying the theory to molecular junctions, we benchmark the extended current expressions against an analytical expression for the electric current and explore the parameter range for which the perturbative electric current expressions are in qualitative agreement. It is found that the perturbative expressions qualitatively agree with the analytical current when the ratio $\frac{{ \Omega}}{\Gamma}$ is small and disagree more severely as $\frac{{ \Omega}}{\Gamma}$ increases. The current characteristics in a equilibrium description are then compared to non-equilibrium descriptions. It is found that terms in the extended current expression do not vanish in the non-equilibrium description and, as a result, we observed significantly different current characteristics such as constructive interference in the conductance characteristics relative to the destructive interference predicted by equilibrium descriptions. In addition, we used the effective temperature to show that the probability distribution of finding the classical particle could be approximated with a Boltzmann-like probability distribution. In the final section, we showed that the non-equilibrium description of nuclear motion can give rise to the Landauer blowtorch effect via the emergence of effective adiabatic potentials with multiple minima. The particular example that we considered consists of an effective potential that varies from a single to a triple potential with voltage and we explored the impacts that the Landauer blowtorch effect has on the trajectories, current, waiting times and Fano factor.

\begin{center}
{\bf DATA AVAILABILITY}
\end{center}

The data that supports the findings of this study are available within the article.

\clearpage

\clearpage
\appendix
\section{Exact results for current}
We begin with the general expression for time-dependent current
\begin{equation}
\mathcal J_\alpha(t)= 
2 e  \text{Re} \int dt' [  G^<(t, t') \Sigma^A_{\alpha}(t',t)+  G^R(t, t')  \Sigma^<_{\alpha}(t',t) \Big].
\label{Jt}
\end{equation}
Next, we employ wide-band approximation,  transform self-energies to the energy domain whilst leaving molecular Green's functions time-dependent

\begin{widetext}

\begin{equation}
\mathcal J_\alpha(t) = 
2 \text{ Im}  \int_{-\infty}^{+\infty} \frac{d \omega}{2 \pi}   \int_{-\infty}^{+\infty} dt'  e^{-i \omega (t'-t)}  
\Big[ \frac{1}{2} G^<(t,t') \Gamma_\alpha -  G^R(t,t') f_\alpha(\omega) \Gamma_\alpha \big].
\end{equation}
The current should satisfy the continuity equation at each time moment $t$
\begin{equation}
\frac{d N}{dt} = \mathcal J_L(t) + \mathcal J_R(t),
\end{equation}
where $N$ is the total number of electrons in the molecule at  time $t$. Then, using the continuity equation 
we can write left current as
\begin{equation}
\mathcal J_L(t) = (1- \frac{\Gamma_L}{\Gamma}) \mathcal  J_L(t) +  \frac{\Gamma_L}{\Gamma}(N(t) - J_R(t)),
\end{equation}
which yields
\begin{equation}
\mathcal J_L(t) = \frac{\Gamma_L}{\Gamma}  \frac{d N}{dt} 
 -  2 \frac{\Gamma_L \Gamma_R}{\Gamma}     \text{ Im }  \int_{-\infty}^{+\infty} \frac{d \omega}{2 \pi}   \int_{-\infty}^{+\infty} dt'  e^{-i \omega (t'-t)} G^R(t,t') [f_L(\omega) - f_R(\omega)] 
\end{equation}
Let us  average the above equation over the Langevin trajectory. The average 
\begin{equation}
\Big\langle  \frac{d N}{dt}  \Big\rangle_x =0 
\end{equation}
obviously disappears, otherwise the molecule will accumulate or loose charge continuously.  The averaged current becomes
\begin{equation}
 \mathcal J_L  =
 -  2 \frac{\Gamma_L \Gamma_R}{\Gamma}   \text{ Im }  \int_{-\infty}^{+\infty} \frac{d \omega}{2 \pi}   \int_{-\infty}^{+\infty} dt'  e^{-i \omega (t'-t)}  \langle G^R(t,t') \rangle_x  [f_L(\omega) - f_R(\omega)].
 \label{JL-av}
\end{equation}
Notice, that the averaged electric current does not depend on time, since the retarded Green's function will depend on relative time only once averaged over the stochastic realisations.

Our next goal is to find the explicit expression for the retarded Green's function to enter it into the expression to electric current. We begin with the  equation of motion  for the retarded Green's function
\begin{equation}
(i \partial_t - h(t))G^R(t,t') 
- \int_{-\infty}^{+\infty} dt_1 \Sigma^R(t,t_1) G^R(t_1,t') = \delta(t-t'),
\end{equation}
it is reduced in the wide band approximation to
\begin{equation}
(i \partial_t - h(t))G^R(t,t') -\frac{i}{2} \Gamma G^R(t_1,t') = \delta(t-t').
\end{equation}
This differential equation can be resolved analytically and yields
\begin{equation}
G^R(t,t') = -i \theta(t-t') e^{-\frac{1}{2} \Gamma (t-t')} e^{- i \int^t_{t'} dt_1 h(t_1)}
\label{gr}
\end{equation}
Substituting retarded Green's function (\ref{gr}) into the expression for current (\ref{JL-av}) gives
\begin{equation}
\mathcal  J_L  =
   2 \frac{\Gamma_L \Gamma_R}{\Gamma}   \text{ Re }  \int_{-\infty}^{+\infty} \frac{d \omega}{2 \pi}   \int_{-\infty}^{t} dt'  e^{-i (\omega -\frac{i}{2} \Gamma) (t-t')}  \Big\langle e^{- i \int^t_{t'} dt_1 h(t_1)}  \Big\rangle_x [f_L(\omega) - f_R(\omega)].
 \label{JL-av1}
\end{equation}
Changing variables of integration to $\tau=t-t'$ and taking into account that quantity ${{ } \langle}e^{- i \int^t_{t'} dt_1 h(t_1)}  \rangle_x$ depends on relative time only once averages over Langevin trajectory we arrive to 
\begin{equation}
\mathcal  J_L =
   2 \frac{\Gamma_L \Gamma_R}{\Gamma}   \text{ Re }  \int_{-\infty}^{+\infty} \frac{d \omega}{2 \pi}   \int^{\infty}_{0} d\tau  e^{-i (\omega -\frac{i}{2} \Gamma)\tau}  \Big\langle e^{- i \int^\tau_{0} dt_1 h(t_1)}  \Big\rangle_x [f_L(\omega) - f_R(\omega)].
 \label{JL-av2}
\end{equation}
\end{widetext}

\section{Numerical Algorithm Description}
\label{Numerical Algorithm Description}
This section gives an outline of the technical details of the numerical algorithm used in section \ref{Applications}. 

\

The integration of the Langevin equation in the absence of the fluctuation-dissipation theorem and with position-dependent viscosity and diffusion coefficient requires special care. We used a recently-proposed algorithm to integrate the Langevin equation and produce stochastic trajectory $(x(t), p(t))$ \cite{Sachs2017}. This process is repeated $n$ times to produce a total of $n$ Langevin trajectories. 

\

The simulations involved calculating various averaged quantities and we briefly provide some more information. Sections \ref{Electric Current Benchmarking}, \ref{Dynamical Blockade} and \ref{Blowtorch Effect} involved averaging various dynamical quantities (electric current $\mathcal{J}(x, p)$ and effective temperature $T(x)$, for example) over nuclear motion at specific voltages. This was easily achieved by running many Langevin trajectories in the long-time limit and then averaging all the resultant trajectories. Mathematically speaking, this means that we numerically performed the calculation: 
\begin{equation}
\langle f \rangle = \sum_{i}^{n} \lim_{T \rightarrow \infty} \frac{1}{T} \int^{t_{0} + T}_{t_{0}} dt f_{i}(t),
\end{equation}
for a given dynamical variable $f(x, p)$ as evaluated with $n$ Langevin trajectories. The frequency distributions $P_{neq}(x)$ were calculated by dividing the position space into bins and running many trajectories from an initial starting point. A probability density distribution of finding the particle at bin $x$ was then constructed from the Langevin trajectories which, in turn, was used to construct the distribution of $P_{neq}(x)$. Simulations were performed using $500$ Langevin trajectories, each with a simulation time of $2 \times 10^7$ time steps; results were found to be independent of the choice of initial condition.

\

Since the acceleration is not available directly from the used integration algorithm, the acceleration-dependent term in (\ref{transmission_corrections}) is computed as 
\begin{equation}
\langle \ddot x  B( x, \omega) \rangle_x = \langle \frac{1}{m} \Big( - \frac{\partial U}{\partial x} + F(x) - \dot x \xi(x) \Big) B x, \omega) \rangle_x,
\end{equation}
where the stochastic force is omitted from the average since it will give a zero contribution to the average for delta-correlated noise.

\clearpage

\end{document}